\documentclass[aps,pra,twocolumn,superscriptaddress,10pt,showkeys]{revtex4-2}

\usepackage{graphicx,xcolor}
\usepackage{amssymb,amsmath,amsthm}
\usepackage{amsfonts,microtype,mathrsfs,bm,bbm,mathtools}
\usepackage{revsymb}
\usepackage{braket}
\usepackage[title,titletoc]{appendix}
\usepackage[english]{babel}

\usepackage{thmtools}
\usepackage[bookmarksnumbered,linktocpage,hypertexnames=false,colorlinks=true,linkcolor=blue,urlcolor=blue,citecolor=blue,anchorcolor=green,breaklinks=true,pdfusetitle]{hyperref}
\usepackage[capitalize,nameinlink]{cleveref}

\newtheorem{thm}{Theorem}
\newtheorem*{thm*}{Theorem}
\newtheorem{lem}[thm]{Lemma}
\newtheorem{prp}[thm]{Proposition}
\newtheorem{cor}[thm]{Corollary}

\crefname{appendix}{Appendix}{Appendices}
\crefname{thm}{Theorem}{Theorems}
\crefname{lem}{Lemma}{Lemmas}
\crefname{prp}{Proposition}{Propositions}
\crefname{cor}{Corollary}{Corollaries}

\DeclareMathOperator{\supp}{supp}
\DeclareMathOperator{\Tr}{Tr}

\DeclareMathOperator{\iden}{id}
\DeclareMathOperator{\rnk}{rank}

\newcommand{\id}{\mathbbm{1}}

\begin{document}

\title{Incompressible quantum information}

\author{Lasse H. Wolff}
\email{lhw@math.ku.dk}
\affiliation{Department of Mathematical Sciences, University of Copenhagen, Universitetsparken 5, 2100 Denmark.}

\author{Matthias Christandl}
\email{christandl@math.ku.dk}
\affiliation{Department of Mathematical Sciences, University of Copenhagen, Universitetsparken 5, 2100 Denmark.}
\affiliation{Microsoft, Redmond, WA 98052, USA.}

\author{Bergfinnur Durhuus}
\email{durhuus@math.ku.dk}
\affiliation{Department of Mathematical Sciences, University of Copenhagen, Universitetsparken 5, 2100 Denmark.}

\author{Christoph Hirche}
\email{hirche@tnt.uni-hannover.de}
\affiliation{Institute for Information Processing (tnt/L3S), Leibniz Universit\"{a}t Hannover, Germany}

\author{Andreas Winter}
\email{andreas.winter@uni-koeln.de}
\affiliation{Department Mathematik/Informatik---Abteilung Informatik, Universit\"{a}t zu K\"{o}ln, Albertus-Magnus-Platz, 50923 K\"{o}ln, Germany}
\affiliation{ICREA {\&} Grup d'Informaci\'{o} Qu\`{a}ntica, Departament de F\'{i}sica, Universitat Aut\`{o}noma de Barcelona, 08193 Bellaterra (Barcelona), Spain}

\date{\today}

\begin{abstract}
For a small classical system correlated with a large one, it is always possible to compress the large system down to a size only dependent on the size of the smaller one, while preserving all correlations up to a given controllable error. Here, we show that such a statement does not hold at all when the large system is quantum mechanical. More precisely, we show that there are bipartite states such that any compression of the large system, by a non-trivial relative factor, will necessarily lead to the destruction of a significant amount of correlation, measured in mutual information. Our results therefore identify a limitation of information-bottleneck approaches to quantum machine learning, showing that a small system specifying the relevant features contained in a large quantum data set does not guarantee a compact encoding of the latter that preserves the relevant correlations.
\end{abstract}

\keywords{quantum information compression, non-classical features, quantum information bottleneck, quantum machine learning, quantum locking}

\maketitle

\section{Introduction}
\label{sec:introduction}
In both classical and quantum information theory, relevant quantities often include auxiliary random variables that have to be optimized over, thus rendering the evaluation of the quantities in question potentially problematic, in particular when little is known regarding the structure of those variables. Most crucially, one might a priori not even know whether the dimension (alphabet size) of the auxiliary random variables can be bounded and therefore whether the optimal value is attainable. When the auxiliary random variables are classical, a number of tools are available that provide the required dimension bounds. The most widely used one being Caratheodory's 
theorem~\cite{caratheodory1907}, but alternative techniques have also been developed~\cite{GohariAnantharam}. In addition, further dimension reduction is possible by directly compressing the auxiliary variables with a classical channel without significantly changing the structural correlations in the relevant distribution~\cite{Hirche2020-np}. Such dimension bounds have allowed important advances in optimizing neural network and deep learning tasks \cite{Tishby2015-gh, Saxe2019-xg, Shwartz-Ziv2017-hw, Geiger2022-lc} and other classical information tasks \cite{Hsu2006-vt, Slonim2000-nq, Shah2019-tg}. The fact that compression of classical auxiliary variables is always possible can even be seen as a theoretical underpinning of why machine learning should, at least in principle, always be able to work efficiently.

Availability of analogous results on compression of auxiliary variables in the quantum setting would similarly pave the way for efficient computation of central quantities in quantum information theory, such as the \emph{quantum information bottleneck function} \cite{Grimsmo2016-vc, Datta2019-bq, Salek2019-bz}, the \emph{squashed entanglement} \cite{Christandl2004-rv, Koashi2004-oj, Brandao2011-gw, ahmed2026squashedentanglement}, the \emph{conditional entanglement of mutual information (CEMI)} \cite{Yang2008-im} and the \emph{quantum capacity with symmetric side channels} \cite{Smith2008-vb, Brandao2012-fg, Smith2008-nz}. The quantum information bottleneck has also been proposed as a training objective for quantum neural networks \cite{CatliWiebe2024QIB}. However, despite the evident importance of this problem, there is a remarkable lack of literature discussing it (see e.g. \cite{Beigi2014-am, Hirche2020-np} for some exceptions).

In the present Letter, we show that the above-mentioned problem of approximate compression has a negative solution, i.e. the natural analogue of the classical result does not hold in the quantum setting. More precisely, it is shown that in general it is not possible to compress auxiliary quantum variables by any non-trivial relative factor without destroying a significant amount of correlation. A precise formulation of the result is provided in \cref{sec:preliminaries} after introducing the necessary notation and definitions. The proof is given in \cref{sec:proof main results}, with some details deferred to appendices. We finish with an outlook and conclusions in \cref{sec:outlook}. 
We also relate the quantum information bottleneck of a binary pure-state source to squashed entanglement; see \cref{app:bottleneck squashed}. Together with~\cite{ahmed2026squashedentanglement}, this gives a binary qubit source whose bottleneck optimum cannot be attained with any finite-dimensional compressed system.

\section{Preliminaries and main results} \label{sec:preliminaries}
In the following, quantum systems and the associated Hilbert spaces will be denoted by the same capital letters $A, B , \ldots$, and their dimensions $|A|, |B|, \ldots$ are assumed to be finite. Similarly, the composite quantum system with constituent systems $A, B, C, \ldots $ and the associated Hilbert space $A\otimes B\otimes C\otimes\cdots$ will be abbreviated as $ABC \cdots$. Furthermore, for an operator $X \in \mathcal{B}(A)$ on a system $A$, $\| X \|_p, p\geq 1$, will denote the usual Schatten norm of $X$ i.e., $\| X \|_p = (\sum_i | s_i |^p)^{1/p}$ where $s_i$ is the $i^\text{th}$ singular value of $X$. An operator $\rho \in \mathcal{B}(A)$ is a quantum state on $A$ if it is positive semi-definite $\rho \geq 0$ and has trace $1$. The von Neumann entropy of a state with orthonormal eigen-decomposition $\rho = \sum_i \lambda_i \ket{e_i}\!\bra{e_i}$ is given by $H (\rho) \coloneqq - \Tr \rho \log  \rho  = - \sum_i \lambda_i \log \lambda_i$. For a multi-partite state $\rho \in \mathcal{B}(ABC)$, we denote the reduced marginal states by $\rho_B \coloneqq \Tr_{AC} \rho$ and so on, and we abbreviate the entropies of the reduced states as $ H (A)_\rho \coloneqq H (\rho_{A})$. Other entropic functions of particular interest include the \emph{mutual information} $I(A:B)_\rho \coloneqq H (A)_\rho + H (B)_\rho - H (AB)_\rho$ and the \emph{conditional entropy} $ H (A|B)_\rho \coloneqq H (AB)_\rho - H (B)_\rho$. For two systems $A , B$, a quantum channel $T$ from $A$ to $B$ is a linear map $T : \mathcal{B}(A) \rightarrow \mathcal{B}(B)$ that is completely positive and trace-preserving. For any system $A$, $\iden_A : \mathcal{B}(A) \rightarrow \mathcal{B}(A)$ and $\id_A : A \rightarrow A$, respectively, denote the identity quantum channel and the identity operator on $A$. Throughout, $\Tr$ denotes the un-normalised trace and all logarithms are base two.

We consider a state $\rho$ of a bi-partite system $AB$, where $B$ is thought of as representing an auxiliary quantum variable. Taking mutual information as a measure of correlation between the systems and representing compression by a quantum channel, a precise formulation of the central question described in the Introduction is whether, for any such $\rho$ and $\epsilon>0$, there exists a quantum channel $T: \mathcal{B}(B) \rightarrow \mathcal{B}(C)$, with $|C|$ depending only on $|A|$ and $\epsilon$, such that   
\begin{equation}\label{eq: I-bound}I(A:B)_\rho - I(A:C)_{(\iden_A \otimes T )( \rho )} \leq \epsilon\,.
\end{equation}
The announced negative answer to this question is contained in the following main result of this Letter. 
\begin{thm} 
\label{thm:main thm}
There is a function $f(x) = 1 + O( x^{- 1 / 2 })$, $f(x) > 2.4 \times 10^{-4}$, such that for any non-trivial quantum system $A$ there exists, for all sufficiently large systems $B$ (compared with $A$), a state $\rho \in \mathcal{B}(AB)$ with the property that for any quantum channel $T : \mathcal{B}(B) \rightarrow \mathcal{B}(C)$ with $|C| \leq |B|$, we have 
\begin{equation} 
\label{eq:eq main thm}
\begin{aligned}
I(A:B)_\rho &- I(A:C)_{(\iden_A \otimes T )( \rho )}\\
&\geq \frac{f(|A|)\log e}{4} \left( 1 - \frac{|C|}{|B|} \right)^2 .
\end{aligned}
\end{equation}
\end{thm}

\cref{thm:main thm} is proved in \cref{sec:proof main results} (for $|A|\geq 4$), where one can also find a choice of the function $f(|A|)$. As the proof shows, states satisfying \cref{eq:eq main thm} are not rare, but belong to a large family of randomly constructed states. The cases $|A|=2,3$ are treated in \cref{app:extension to d=2 and 3 cases} due to some added complications. If $|A| \geq 3$, the state $\rho$ can even be assumed classical on $A$ if one accepts a slightly worse prefactor in the lower bound of \cref{eq:eq main thm}, see \cref{app:extension to classical A}. It remains open whether the compression can be ruled out if the $A$ system is classical with dimension $2$, although there is some evidence to the contrary, as it follows from \emph{Jordan's Lemma} that compression is actually possible in this restricted setting if one further assumes the conditional quantum states on $B$ to be proportional to projectors (see \cref{cor:compression of 2D projectors} in \cref{app:KI compression}).

While \cref{thm:main thm} rules out any form of compression of auxiliary variables with arbitrarily small loss of mutual information, the stated lower bound can be improved significantly when the system $A$ is large enough, as shown by the following result. It follows from \cref{thm:main thm} by considering tensor products of the states exhibited there, with the proof given in \cref{app:tensor corollary proof}: 
\begin{cor} 
\label{cor:tensor prod construction mutual info}
For any sufficiently large quantum system $A$ and any number $D \geq 1$, there exists a state $\rho \in \mathcal{B} ( AB )$ on $A$ composed with an arbitrary sufficiently large system $B$ (compared to $A$ and $D$), such that for all quantum channels $T : \mathcal{B}(B) \rightarrow \mathcal{B}(C)$ with $ |C| \leq D$, we have
\begin{equation*}
I(A:B)_{\rho} - I(A:C)_{ (\iden_A \otimes T) (\rho) } \geq 0.03498 \times \log  |A|  .
\end{equation*}
\end{cor}
A construction with random states can be shown to result in a strengthened information loss in \cref{cor:tensor prod construction mutual info}, reaching the essentially optimal value $2\log|A|-O(\log\log|A|)$, even if an unlimited finite classical record is retained alongside the compressed quantum system. A version with classical reference system $A$ results similarly in an essentially optimal information loss of $\log|A|-O(\log\log|A|)$. The full statements and proofs are given in \cref{app:random states}.

The results above are surprising considering that the strong compression described earlier is possible classically. Specifically, as shown in \cite{Hirche2020-np}, for any joint probability distribution $p_{XY}$, we can always find a new variable $Z$, of alphabet size bounded by $|Z|\leq(4/\epsilon)^{|X|}$, and a classical compression channel $T:Y\rightarrow Z$, with trace-distance recovery error at most $\epsilon$ and mutual-information loss at most $\epsilon\log|X|+g(\epsilon)$, where $g(x)\coloneqq(1+x)\log(1+x)-x\log x$. A similar result holds for quantum states $\rho_{AB}$ that are classical on the system $B$ that we want to compress, with compressed dimension bounded by $(1+4/\epsilon)^{2|A|^2}$ at the same recovery error. For the benefit of the reader, self-contained proofs with sharper bounds are provided in \cref{app:KI classical cases}.
Generalizing these results even further, the aforementioned compression is possible as long as the auxiliary variables have Koashi-Imoto quantum parts bounded independently of $|B|$. The compression and recovery scheme and dimension bound are given in \cref{app:KI compression}.

Under the same restriction on the quantum parts, classical support reduction gives exact auxiliary bounds and attainment for the quantum information bottleneck and squashed-entanglement optimizations; see \cref{app:restricted optimizations}. However, for a binary source as considered in \cref{app:bottleneck squashed}, approaching the optimum requires unbounded quantum parts even with unrestricted finite classical storage.

\section{Proofs of the main results} \label{sec:proof main results}
As a prerequisite for the proof of \cref{thm:main thm}, we shall first establish an equivalence between the question of compressibility of the auxiliary quantum system at a small cost in mutual information, as formulated in the previous section, and the ability to recover the original state from the compressed state, up to a small deviation in trace norm, by applying a \emph{recovery channel}. This is the content of the following lemma.
\begin{lem} \label{lemma:hypothesis are equivalent}
Let $\rho \in \mathcal{B}(AB)$ be an arbitrary bipartite quantum state. Suppose that $T : \mathcal{B}(B) \rightarrow \mathcal{B}(C)$ is a quantum channel satisfying \cref{eq: I-bound}. Then, we can construct a recovery channel $R : \mathcal{B}(C) \rightarrow \mathcal{B}(B)$ such that
\begin{equation} \label{eq:lemma equivalence:result1}
\left\| \vphantom{3^3} \rho - \iden_A \otimes ( R \circ T) (\rho) \right\|_1 \leq 2 \sqrt{\frac{\epsilon}{\log e}} .
\end{equation}
Conversely, let $T : \mathcal{B}(B) \rightarrow \mathcal{B}(C)$ and $R : \mathcal{B}(C) \rightarrow \mathcal{B}(B) $ be two quantum channels such that $ \|  \rho - \iden_A \otimes ( R \circ T) (\rho) \|_1 \leq \delta$. Then, $T$ satisfies \cref{eq: I-bound} with $\epsilon = \delta \log |A| + g (\delta / 2 )$.
\end{lem}
The first statement of the lemma follows more or less straightforwardly from a recovery result for approximate quantum Markov chains \cite{Fawzi2015-yy} (or one of its many refinements \cite{Junge2018-ib}), while the second statement follows from the Alicki-Fannes inequality for the conditional entropy (in the form from \cite{Winter2016_continuitybounds}). For the benefit of the reader, \cref{lemma:hypothesis are equivalent} is proved rigorously in \cref{app:proof lemma equivalence}. 

To prove \cref{thm:main thm}, we shall construct bipartite states for which we cannot locally compress and recover the original state with high trace-norm precision, which by \cref{lemma:hypothesis are equivalent} suffices to prove the theorem. The states are constructed from random unitaries that form a \emph{quantum expander}, a result due to Hastings \cite{Hastings2007-ev}. In the proof, the unitaries of the state appear together under a trace with the Kraus operators of the postulated compression-decompression map. The quantum expander property guarantees sufficient scrambling of the Kraus operators, which must have small rank and overlap with the identity if significant compression have taken place, leading to the desired bound. \cref{prop:main counterexample} below deals with the case $|A| \geq 4$. For the sake of clarity of the presentation, the cases $|A| = 2 , 3$, that require special treatment, are relegated to \cref{app:extension to d=2 and 3 cases}. 

\begin{prp}
\label{prop:main counterexample}
Let $A$ be any quantum system of dimension $|A| \geq 4$ and define the number $K \coloneqq \lfloor |A| / 2 \rfloor$. Let further $B$ be an arbitrary additional quantum system and $U_1 , \ldots , U_K \in \mathcal{U} (B)$ be $K$ Haar-random unitaries on $B$. Define now the operator $\rho_{AB} \in \mathcal{B}(AB)$ as
\begin{equation} 
\label{eq:prop d geq 4: rho_AB def}
\begin{aligned}
\rho_{AB} = \frac{1}{ 2 K |B|} \sum_{j=1}^{K} \left( \vphantom{\frac{1}{1}} \big( \ket{2j-1}\!\bra{2j-1}  + \ket{2j}\!\bra{2j}  \big) \otimes \id_B \right. \\
\left. \vphantom{\frac{1}{1}} \mathbin{+} \ket{2j-1}\!\bra{2j} \otimes U_j + \ket{2j}\!\bra{2j-1} \otimes U_j^\dagger \right) \!,
\end{aligned}
\end{equation}
where $\{ \ket{j} \}_{1 \leq j \leq |A|}$ is an orthonormal basis for $A$. Then, $\rho_{AB}$ is a quantum state on $AB$, and for any $0<\varepsilon<1-\sqrt{2K-1}/K$, we have with probability converging to $1$ as $|B| \rightarrow \infty$ (with the convergence only depending on $K$ and $\varepsilon$), that for all pairs of quantum channels $T : \mathcal{B}(B) \rightarrow \mathcal{B}(C)$ and $R : \mathcal{B}(C) \rightarrow \mathcal{B}(B) $, where $C$ is a further arbitrary system, the following is satisfied
\begin{equation} 
\label{eq:prop d geq 4: statement}
\begin{split}
 &\left\| \vphantom{2_2^2} \rho_{AB} - \iden_A \otimes ( R \circ T) (  \rho_{AB} ) \right\|_1 \\
  &\phantom{===}
   \geq \left( 1 - \frac{\sqrt{2K-1}}{K} - \varepsilon \right) \left( 1 - \frac{|C|}{|B| } \right)  .
\end{split}
\end{equation} 
\end{prp}
Note that the state can be constructed by choosing a random $j$ and then applying controlled-$U_j$ with control on $\frac{1}{\sqrt{2}}(\ket{2j-1}+\ket{2j})$ and target the maximally mixed state on $B$.

For the state $\rho_{AB}$ constructed in \cref{eq:prop d geq 4: rho_AB def}, $\|\rho_{AB}-\rho_A\otimes\rho_B\|_1\to1$ in probability as $|B|\to\infty$ at fixed $K$, since $|\Tr U_j|/|B|\to0$ \cite{Petz2004-oo}. Thus, in the limits $|C|/|B|\to0$, $K\to\infty$ and $\varepsilon\to0$, the bound \eqref{eq:prop d geq 4: statement} is asymptotically attained by discarding $B$ and preparing $\rho_B$. In this regime it cannot be improved by a constant factor, and at most a factor of $2$ for an alternative state construction, since the trace norm difference is bounded by $2$.

\begin{proof}[Proof of \cref{prop:main counterexample}]\label{proof:of main counterexample}
Let $\rho_{AB}$ be defined as in \cref{eq:prop d geq 4: rho_AB def}. It is easily seen that $\rho_{AB}$ is a quantum state, and we shall now show that it satisfies \cref{eq:prop d geq 4: statement} (for all $R,T$, with probability approaching $1$ as $|B| \rightarrow \infty$). Thus, consider any pair of quantum channels $T : \mathcal{B}(B) \rightarrow \mathcal{B}(C)$ and $R : \mathcal{B}(C) \rightarrow \mathcal{B}(B) $, where $C$ is some arbitrary system. Define now the matrix $H$ as 
\begin{equation*} 
\label{eq:proof main counterexmp:H def}
H \coloneqq \sum_{j=1}^{K} \left(  \ket{2j-1}\!\bra{2j} \otimes U_j +  \ket{2j}\!\bra{2j-1} \otimes U_j^\dagger \right) ,
\end{equation*}
$H$ is Hermitian and satisfies $H^2 = ( \sum_{j=1}^{2K} \ket{j} \bra{j} ) \otimes \id_B$, hence $\| H \|_\infty = 1$. It thus follows from the H{\"o}lder inequality for Schatten norms: $|\Tr (A^\dagger B)| \leq \| A \|_\infty \| B \|_1$ \cite[Eq.~(1.174)]{Watrous2018-sx}, and linearity that
\begin{equation} 
\label{eq:proof main counterexmp:long step 1}
\begin{split}
 &\left\| \vphantom{2^2} \rho_{AB} - \iden_A \otimes ( R \circ T ) (  \rho_{AB} ) \right\|_1 \\
  &\phantom{=}
   \geq \Tr \left[ H \left( \rho_{AB} - \iden_A \otimes ( R \circ T ) (  \rho_{AB} )   \vphantom{2^2} \right) \right] \\
  &\phantom{=}
   = \frac{1}{2 K |B|} \sum_{j=1}^K \Tr \left[ U_j \left(\iden_B - R \circ T \right) ( U_j^\dagger ) \right. \\
  &\phantom{======}
   \left. \quad \qquad + U_j^\dagger \left(\iden_B - R \circ T \right) ( U_j )  \right] .
\end{split}
\end{equation}
The composition $R \circ T : \mathcal{B}(B) \rightarrow \mathcal{B}(B)$ is a quantum channel, meaning that it can be written in Kraus decomposition as $R \circ T = \sum_l K_l \cdot K_l^\dagger$, for some $\{ K_l \}_l \subseteq \mathcal{B}(B)$, satisfying $\sum_l K_l^\dagger K_l = \id_B$. This lets us rewrite 
\begin{equation} \label{eq:proof main counterexmp:long step 2}
\begin{aligned}
\sum_{j=1}^K \Tr \left[ U_j \left(\iden_B - R \circ T \right) ( U_j^\dagger )  + U_j^\dagger \left(\iden_B - R \circ T \right) ( U_j )   \right] \\
= \sum_l \sum_{j=1}^K \left\| \left[ K_l , U_j \vphantom{3_3^2} \right] \right\|_2^2 ,
\end{aligned}
\end{equation}
where we have used $\sum_l \| K_l \|_2^2 = \sum_l\Tr K_lK_l^\dagger = \sum_l\Tr K_l^\dagger K_l = \Tr\id_B = |B|$ and the fact that the $U_j$'s are unitaries. We now use Hastings's result \cite{Hastings2007-ev} stating that for a fixed number of $K \geq 2$ Haar-randomly drawn unitaries $U_1 , \ldots , U_K \in \mathcal{U}(B)$ on any space $B$, the unital self-adjoint linear map $\sum_{j=1}^K ( U_j \cdot U_j^\dagger + U_j^\dagger \cdot U_j ) / (2K)$ will for any $\varepsilon > 0$ have its second largest eigenvalue $\lambda_2$ upper bounded by $\lambda_2 \leq 2 \sqrt{2K-1} / (2K) + \varepsilon$ with probability converging to $1$ as $|B| \rightarrow \infty$. This result can be reformulated into stating that for any $\varepsilon > 0$, we have with probability converging to $1$ as $|B| \rightarrow \infty$ that the following is satisfied for any $X \in \mathcal{B}(B)$:
\begin{equation} 
\label{eq:proof main counterexmp:reformulation hastings}
\begin{split}
\frac{1}{2K} &\sum_{j=1}^K \left\| \left[ X , U_j \vphantom{3_3^2} \right] \right\|_2^2 \\
  &\geq \left( 1 - \frac{\sqrt{2K-1}}{K}-\varepsilon \right) \left( \| X \|_2^2 - \frac{| \Tr X |^2}{|B|} \right) ,
\end{split}
\end{equation}
as proven for completeness by \cref{app:prop:commutator result from spectral gap} in \cref{app:sec:tedious calculations}. We now combine \cref{eq:proof main counterexmp:long step 1} and \cref{eq:proof main counterexmp:long step 2} while using \cref{eq:proof main counterexmp:reformulation hastings} with $X = K_l$ once for every $l$ to obtain:
\begin{equation*} 
\label{eq:proof main counterexmp:long step 3}
\begin{split}
 &\left\| \vphantom{2^2} \rho_{AB} - \iden_A \otimes ( R \circ T ) (  \rho_{AB} ) \right\|_1 \\
  &\phantom{=}
   \geq \frac{1}{|B|} \sum_l \frac{1}{2K} \sum_{j=1}^K \left\| \left[ K_l , U_j \vphantom{3_3^2} \right] \right\|_2^2 \\
  &\phantom{=}
   \geq \left( 1 - \frac{\sqrt{2K-1}}{K}-\varepsilon \right) \frac{1}{|B|} \sum_l \left( \| K_l \|_2^2 - \frac{| \Tr K_l |^2}{|B|} \right) \\
  &\phantom{=}
   \geq \left( 1 - \frac{\sqrt{2K-1}}{K}-\varepsilon \right) \left( 1 - \frac{|C|}{|B|} \right) ,
\end{split}
\end{equation*}
where we have used the Cauchy-Schwarz inequality for the Frobenius norm to get $| \Tr K_l |^2 \leq \| P_{\supp K_l} \|_2^2 \| K_l \|_2^2 = \rnk (K_l) \| K_l \|_2^2 \leq |C| \| K_l \|_2^2$ for all $l$, with $P_{\supp K_l}$ being the projector onto the support of $K_l$, and the fact that $\sum_l \| K_l \|_2^2 = |B|$ to obtain the final inequality above. The inequalities $(\rnk K_l) \leq |C|$ follow from the fact that the Kraus operators of $R \circ T$ can w.l.o.g. be chosen as products of the Kraus operators for the channels $R$ and $T$ respectively, which are themselves $|B| \times |C|$ or $|C| \times |B|$ matrices with rank bounded by $|C|$. 
\end{proof}

We are now finally ready to prove \cref{thm:main thm}.

\begin{proof}[Proof of \cref{thm:main thm}] \label{proof:proof of main thm}
If $|A| \geq 4$, \cref{prop:main counterexample} entails that there must for all sufficiently large $B$ systems exist a state $\rho_{AB}$ satisfying \cref{eq:prop d geq 4: statement} (since it must hold with nonzero probability given the random construction). \cref{prop:d=3 special case} and \cref{prop:d=2 case} gives similar results for respectively the $|A|=3$ and $|A|=2$ case. In conclusion, there exists a strictly positive function $F(|A|) > 0$ such that for any non-trivial $A$, every sufficiently large system $B$ admits a state $\rho_{AB}$ such that for all pairs of channels $T : \mathcal{B}(B) \rightarrow \mathcal{B}(C)$, $R : \mathcal{B}(C) \rightarrow \mathcal{B}(B)$ with $|C|\leq|B|$, we have 
\begin{equation} 
\label{eq:proof main thm:l1}
\left\| \vphantom{2_2^2} \rho_{AB} - \iden_A \otimes ( R \circ T) (  \rho_{AB} ) \right\|_1 \geq F(|A|) \left( 1 - \frac{|C|}{|B|} \right) .
\end{equation}
We can specifically choose: $F(|A|) \coloneqq 1  - \sqrt{2K-1} / K- \varepsilon$, with $K \coloneqq \lfloor |A| / 2 \rfloor$, when $|A| \geq 4$, and choose $F(3) \coloneqq 1/2 - \sqrt{5}/6 - \varepsilon$ and $F(2) \coloneqq0.0156 - \varepsilon$ for all sufficiently small $\varepsilon > 0$. Setting e.g. $\varepsilon=1/(10000 \sqrt{|A|})$ gives $F(|A|)>0$ and $F(|A|)=1+O(1/\sqrt{|A|})$. \cref{lemma:hypothesis are equivalent} now entails
\begin{equation} 
\label{eq:proof main thm:l2}
\begin{aligned}
 I(A:B)_\rho &- I(A:C)_{(\iden_A \otimes T )( \rho )}\\
  &\geq \frac{F(|A|)^2\log e}{4} \left( 1 - \frac{|C|}{|B|}\right)^2 ,
\end{aligned}
\end{equation}
since otherwise, there would by \cref{eq:lemma equivalence:result1} from \cref{lemma:hypothesis are equivalent} exist a recovery map $R$ such that \cref{eq:proof main thm:l1} was contradicted. \cref{eq:proof main thm:l2} now proves \cref{eq:eq main thm} with $f(|A|) \coloneqq F(|A|)^2$. 
\end{proof}

\section{Discussion and outlook} \label{sec:outlook}

We have seen that it is in general not possible to compress the larger system of a bipartite quantum state to the size of the smaller one (or a function of it) without destroying a significant amount of correlation. This is in sharp contrast to the classical situation, where this is possible although the net argument gives an exponential bound (in practice, much better scaling might be observed). `Classical' in fact extends to the situation where the large system to be compressed is classical, while the smaller system can be quantum. The results in this Letter therefore apply to the opposite situation, where it is the truly quantum system we want to compress, while the other one can be classical. 

Cast in a machine learning context, this situation arises when a small amount of conceptual ideas (features) are conveyed to a neural net via a large amount of raw data. The neural net then plays the role of the compressor. A requirement for any sensible compressor is of course the preservation of the essential correlation of its output with the features. Whereas this is always possible in principle for classical data (and even observed in practice due the remarkable performance of neural networks), our work shows that quantum machine learning designed according to the same principle is here facing a fundamental obstruction. Turning the findings on their head, we show that classical features encoded in raw quantum data may be secured from being scraped by a quantum machine learning system with unbounded computing power solely restricted by a quantum memory bound. 

Viewed in this sense, our work can also be seen in the line of work on locking classical information in quantum states \cite{PhysRevLett.92.067902}. Locking constructions also lend themselves to counterexamples of compressibility, although far from the extremes we have encountered in this Letter. The reason is that typically one starts with pure states, which are then locked with few unitaries (indexed by a key system $K$), and the conditional states on $B$ are thus supported on a common subspace of dimension at most $|A||K|$. 

\pagebreak
\section*{Acknowledgements} 
The authors acknowledge the use of ChatGPT 6.0 Astra in proving~\cref{app:lem:coarse pinching}, which provided a missing piece to the tighter argument in~\cref{app:random states}, as well as general proofreading.  
However, no language model was used in proving the main result of this Letter (\cref{thm:main thm}), its corollaries and refinements, and in general everything other than the stated arguments from \cref{app:random states}. All parts of the Letter were written by the authors.

M.C., B.D., and L.H.W. acknowledge financial support from VILLUM FONDEN via the QMATH Centre of Excellence (Grant No.10059). C.H. was supported by the Deutsche Forschungsgemeinschaft (DFG, German Research Foundation), project 550206990; by Germany's Federal Ministry of
Research, Technology and Space (BMFTR) through the QC service center QUICS,
grant 13N17418, and QUANDA, grant 01MQ26002B; and by Germany's Excellence Strategy, EXC 2123/2
QuantumFrontiers, project 390837967. A.W. is supported by the Alexander von Humboldt Foundation; furthermore by the German Research Foundation (DFG) under Germany’s Excellence Strategy: Cluster of Excellence ``Matter and Light for Quantum Computing'' (ML4Q) EXC 2004/2–390534769; and in part by the Spanish MICIN (project PID2022-141283NB-I00) with the support of FEDER funds.

\bibliography{refs}

\onecolumngrid

\clearpage

\begin{centering}
\section*{Appendices}
\end{centering}

\begin{appendices}
\appendix
\crefalias{section}{appendix}
\crefalias{subsection}{appendix}
\crefalias{subsubsection}{appendix}

\section{Proof of \texorpdfstring{\cref{lemma:hypothesis are equivalent}}{cor:tensor product construction}} \label{app:proof lemma equivalence}

We here rigorously prove \cref{lemma:hypothesis are equivalent}, which was vital for the proof of the main result of the Letter. 

\begin{proof}[Proof of \cref{lemma:hypothesis are equivalent}] \label{proof:equivalent hypothesis}
To prove the first statement of the lemma, assume that \cref{eq: I-bound} holds for some arbitrary quantum state $\rho \in \mathcal{B}(AB)$, channel $T : \mathcal{B}(B) \rightarrow \mathcal{B}(C)$ and $\epsilon > 0$, and consider the \emph{Stinespring representation} $T = \Tr_E \Gamma ( \ \cdot \ ) \Gamma^\dagger$ of $T$, with $\Gamma : B \rightarrow C E$ being an isometry (see e.g. \cite[sec. 2]{Watrous2018-sx}). Defining $\omega_{ACE} \coloneqq (\id_A \otimes \Gamma) \rho (\id_A \otimes \Gamma^\dagger)$, it follows that $\omega_A = \Tr_{CE} \omega = \rho_A$, $\omega_{AC} = \Tr_E \omega = (\iden_A \otimes T)(\rho)$, and since $\Gamma$ is an isometry, $\rho_{AB}$ and $\omega_{ACE}$ have the same nonzero spectra, as do $\rho_B$ and $\omega_{CE}$. Using these results and the assumptions above, we get
\begin{equation} 
\label{eq:proof equivalence:s1}
\begin{split}
  \epsilon &\geq I(A:B)_{\rho} - I(A:C)_{(\iden_A \otimes T)(\rho)} = I(A:CE)_\omega - I (A:C)_\omega 
   \eqqcolon I(A:E | C)_\omega .
\end{split}
\end{equation}
It follows from \cref{eq:proof equivalence:s1} and a recovery result for approximate quantum Markov chains \cite{Fawzi2015-yy} that there exists a recovery map $R' : \mathcal{B}(C) \rightarrow \mathcal{B}(CE)$ satisfying
\begin{equation} \label{eq:proof equivalence:s2}
\frac{\log e}{4} \| \omega_{ACE} - \iden_A \otimes R' (\omega_{AC}) \|_1^2 \leq I(A:E | C)_\omega \leq \epsilon .
\end{equation}
(the result from \cite{Fawzi2015-yy} is usually formulated in terms of the fidelity $F ( \rho , \sigma) \coloneqq \| \sqrt{\rho} \sqrt{\sigma} \|_1$, but it entails \cref{eq:proof equivalence:s2} via the inequalities $\| \rho - \sigma \|_1^2 / 4 \leq 1  - F(\rho , \sigma)^2$ \cite[thm. 3.33]{Watrous2018-sx} and $1 - 2^{-x} \leq  x / \log e $). We can now finally define the desired recovery map $R : \mathcal{B}(C) \rightarrow \mathcal{B}(B)$ as:
\begin{equation} 
\label{eq:proof equivalence:s3}
R(X) \coloneqq \Gamma^\dagger R' (X) \Gamma + \left( \Tr \left[ R' (X) (\id_{CE}-\Gamma \Gamma^\dagger) \right]  \right) \sigma_B,
\end{equation}
where $\sigma_B \in \mathcal{B}(B)$ is an arbitrarily chosen state, introduced to make $R$ trace-preserving. Combining \cref{eq:proof equivalence:s3} with \cref{eq:proof equivalence:s2}, while using that $\Gamma$ is an isometry, we get
\begin{equation} 
\label{eq:proof equivalence:s4}
\begin{split}
 \left\| \rho_{AB} - \iden_A \otimes ( R \circ T) (\rho_{AB}) \vphantom{3^3} \right\|_1 
   &= \left\| (\id_A \otimes \Gamma) \left(  \rho_{AB}  - \iden_A \otimes  R (\omega_{AC}) \vphantom{3^3} \right) (\id_A \otimes \Gamma^\dagger) \right\|_1 \\ 
  &= \left\| (\iden_A \otimes S_\Gamma) \left[ \omega_{A C E} -\left( \iden_A \otimes  R' (\omega_{AC}) \right) \right] \right\|_1 \\
  &\leq \left\| \omega_{A C E} -\left( \iden_A \otimes  R' (\omega_{AC}) \right)  \right\|_1 
   \leq 2 \sqrt{\frac{\epsilon}{\log e}} ,
\end{split}
\end{equation} 
where we note that $\Gamma \Gamma^\dagger $ is a projector, and we have defined the quantum channel $S_\Gamma : \mathcal{B}(CE) \rightarrow \mathcal{B}(CE)$ as $S_\Gamma (X) \coloneqq \Gamma \Gamma^\dagger X \Gamma \Gamma^\dagger + \Tr [X(\id_{CE} - \Gamma \Gamma^\dagger)] \Gamma \sigma_B \Gamma^\dagger$. Note also that $(\iden_A \otimes S_\Gamma) (\omega_{ACE}) = \omega_{ACE}$ follows from $\omega_{ACE} = (\id_A \otimes \Gamma \Gamma^\dagger) \omega_{ACE} (\id_A \otimes \Gamma \Gamma^\dagger)$. The final inequalities in \cref{eq:proof equivalence:s4} follow from \cref{eq:proof equivalence:s2} and the monotonicity of the trace norm under quantum channels.

To prove the second statement of the lemma, assume that $\| \rho_{AB} - \iden_A \otimes ( R \circ T) (\rho_{AB}) \|_1 \leq \delta$ holds for some quantum state $\rho \in \mathcal{B}(AB)$ and channels $T : \mathcal{B}(B) \rightarrow \mathcal{B}(C)$, $R : \mathcal{B}(C) \rightarrow \mathcal{B}(B)$. It then follows from $\rho_{AB}$ and $(\iden_A \otimes (R \circ T))(\rho_{AB})$ having the same marginal $\rho_A$ on the $A$ system and the Alicki-Fannes inequality for the conditional entropy in the formulation of \cite{Winter2016_continuitybounds}, that
\begin{equation} 
\label{eq:proof equivalence:second s1}
\begin{split}
 I(A:B)_{\rho} - I(A:B)_{(\iden_A \otimes (R \circ T) ) \rho}  
= H(A|B)_{(\iden_A \otimes ( R \circ T)) \rho} - H(A|B)_\rho \leq \delta \log |A| + g \left(  \delta / 2 \right) ,
\end{split}
\end{equation}
where $ g (x) = (1+x) h_2( x / (1+x))$, with $h_2$ being the binary entropy function. Meanwhile, it follows from the data-processing inequality for the mutual information that we have $I(A:B)_{(\iden_A \otimes (R \circ T) ) \rho} \leq I(A:C)_{(\iden_A \otimes T )( \rho)}$, which when plugged into \cref{eq:proof equivalence:second s1} shows that \cref{eq: I-bound} is satisfied with $\epsilon = \delta \log |A| + g (\delta / 2 )$.
\end{proof}

\section{Proof of \texorpdfstring{\cref{cor:tensor prod construction mutual info}}{cor:tensor product construction}} \label{app:tensor corollary proof}

We here prove \cref{cor:tensor prod construction mutual info}, which follows from \cref{thm:main thm} by taking tensor products of the state from the theorem. 

\begin{proof}[Proof of \cref{cor:tensor prod construction mutual info}] \label{proof:cor:tensor product extension}
Consider a bi-partite quantum system $AB$. Assume $|A| \geq 4$ and let $c \in \mathbb{N}$ be an integer in the interval $ 4 \leq c \leq |A|$. Define further $L \coloneqq \lfloor \log |A| / \log c \rfloor$, such that $c^{L} \leq |A| < c^{L+1}$. For each $1 \leq k \leq L$, let $A_k$ be a space of dimension $| A_k | = c$, and let $B_k$ be a further space of dimension $| B_k | = n$, where $n>|A|D/c$ will be chosen sufficiently large below. Thus, $\otimes_{k=1}^L A_k$ is isomorphic to a subspace of $A$. We now choose $B$ sufficiently large such that $\otimes_{k=1}^L B_k \subseteq B$, and the bipartite state $\rho_{AB}$ on $AB$ as the product state:
\begin{equation} \label{eq:proof tensor constrct:state def}
\rho_{AB} \coloneqq \bigotimes_{k=1}^L \Tilde{\rho}_{A_k B_k} ,
\end{equation}
where for each $k$, the quantum state $\Tilde{\rho}_{A_k B_k} \in \mathcal{B}(A_k B_k)$ is constructed like the state $\rho_{AB}$ from \cref{prop:main counterexample}. Thus, the probability that each $\Tilde{\rho}_{A_k B_k}$ satisfies \cref{eq:prop d geq 4: statement} (with the relevant systems) will converge to $1$ as $n \rightarrow \infty$. 

Consider now any quantum channel $T : \mathcal{B}(B) \rightarrow \mathcal{B}(C)$ for some space $C$ with dimension bounded by $|C| \leq D$. Since $\rho_{AB} \in \mathcal{B}(A_1 \ldots A_L B_1 \ldots B_L)$ is by definition \eqref{eq:proof tensor constrct:state def} a product state between the $A_k B_k$ systems, it follows that
\begin{equation} \label{eq:prooftensor:s2}
\begin{aligned}
I( A : B )_{\rho} = \sum_{k=1}^L I(A_k : B_k)_{\tilde{\rho}_{A_k B_k}} .
\end{aligned}
\end{equation}
Furthermore, for any quantum systems $Y , X_1 , \ldots , X_L$, it follows from adding and subtracting terms that:
\begin{equation} \label{eq:prooftensor:s1}
\begin{aligned}
I(X_1  \ldots  X_L : Y) = 
H \left(X_1 \ldots X_L \right) - \sum_{j=1}^L H(X_j) + \sum_{j=1}^L I (X_j : Y X_1 \ldots X_{j-1}) ,
\end{aligned}
\end{equation}
with $I (X_j : Y X_1 \ldots X_{j-1}) \coloneqq I(X_1 : Y)$ in case $j=1$. Thus, applying \cref{eq:prooftensor:s1} to the output state $(\iden_A \otimes T) \rho_{AB} \in \mathcal{B}(A_1 \cdots A_L C)$, whose reduced state on $A_1 \ldots A_L$ is a product state, we get
\begin{equation} \label{eq:prooftensor:s3}
\begin{aligned}
I(A:C)_{(\iden_A \otimes T) \rho_{AB}} =  I(A_1 \ldots A_L:C)_{(\iden_A \otimes T) \rho_{AB}} = \sum_{k=1}^L I (A_k : C A_1 \cdots A_{k-1})_{(\iden_A \otimes T) \rho_{AB}} ,
\end{aligned}
\end{equation}
since $H( A_1 \ldots A_L)_{(\iden_A \otimes T) \rho_{AB}} = \sum_{k=1}^L H(A_k)_{(\iden_A \otimes T) \rho_{AB}}$.
Then, combining \cref{eq:prooftensor:s2} and \cref{eq:prooftensor:s3} gives us
\begin{equation} 
\label{eq:prooftensor:s4}
\begin{split}
I(A:B)_{\rho_{AB}} - I(A:C)_{(\iden_A \otimes T) \rho_{AB}}
  &= \sum_{k=1}^L \left( I(A_k : B_k)_{\tilde{\rho}_{A_k B_k}}  - I (A_k : C A_1 \cdots A_{k-1})_{(\iden_A \otimes T) \rho_{AB}} \right) \\
  &= \sum_{k=1}^L \left( I(A_k : B_k)_{\tilde{\rho}_{A_k B_k}} - I (A_k : C A_1 \cdots A_{k-1})_{(\iden_{A_k} \otimes \Tilde{T}_k) \tilde{\rho}_{A_k B_k}} \right), 
\end{split}
\end{equation}
where we have for every $1 \leq k \leq L$ defined the modified quantum channel $\Tilde{T}_k : \mathcal{B}(B_k) \rightarrow \mathcal{B}(C A_1 \cdots A_{k-1})$ as follows:
\begin{equation*}
\begin{aligned}
\Tilde{T}_k&\coloneqq\Tr_{A_{k+1}\ldots A_L}\circ
\left(\iden_{\bigotimes_{j\ne k}A_j}\otimes T\right)\circ M_k,\\
\text{where } M_k&:\mathcal B(B_k)\rightarrow
\mathcal B\left(\left(\bigotimes\nolimits_{j\ne k}A_j\right)B\right) \text{ is given by } 
M_k(X) \coloneqq X\otimes\left(\bigotimes\nolimits_{j\ne k}\Tilde\rho_{A_jB_j}\right),
\end{aligned}
\end{equation*}
which has the property that $(\iden_{A_k}\otimes\Tilde T_k)\tilde\rho_{A_kB_k}$ equals the $A_k C A_1\cdots A_{k-1}$ marginal of $(\iden_A\otimes T)\rho_{AB}$ for any $k$, thus proving \cref{eq:prooftensor:s4}. We can now apply \cref{prop:main counterexample} and \cref{lemma:hypothesis are equivalent} to each of the elements in the final sum in \cref{eq:prooftensor:s4} (where the systems $A , B , C$ from \cref{prop:main counterexample} have been replaced with $A_k , B_k , C A_1 \ldots A_{k-1}$) and get for any $0<\varepsilon<1-\sqrt{2\lfloor c/2\rfloor-1}/\lfloor c/2\rfloor$ that the following is satisfied with high probability as $n \rightarrow \infty$.
\begin{equation*}
\label{eq:prooftensor:s5}
\begin{split}
I(A:B)_{\rho} - I(A:C)_{(\iden_A \otimes T) \rho} 
  &\geq \sum_{k=1}^L \frac{\log e}{4} F ( |A_k| )^2 \left( 1  - \frac{| C A_1 \ldots A_{k-1} |}{|B_k|} \right)^2  
\geq \frac{L\log e}{4} \left( 1 - \frac{ D c^{L-1}}{n } \right)^2 F(c)^2  \\
  &\geq \frac{\log|A|\log e}{4\log c} \left( 1 - \frac{\log c}{\log |A|} \right) \left(1-\frac{|A| D}{cn}\right)^2 \left( 1 - \frac{\sqrt{2 \lfloor c / 2 \rfloor - 1 } }{\lfloor c / 2 \rfloor} - \varepsilon \right)^2 \!,
\end{split}
\end{equation*}
with $F(x) \coloneqq 1 - \sqrt{2 \lfloor x / 2 \rfloor -1} / \lfloor x / 2 \rfloor - \varepsilon$. For $c=138$, $(1-\sqrt{137}/69)^2\log e/(4\log138)>0.03498$. Fixing $\varepsilon>0$ sufficiently small, then $|A|$ sufficiently large and finally a finite $n$ sufficiently large, thus proves the corollary for every $B$ with $|B|\geq n^L$. 
\end{proof}

\section{Relating quantum expanders to bounds on commutator sums} \label{app:sec:tedious calculations}

We here rigorously prove the following result, relating quantum expander properties to lower bounds on commutator sums, and which is used in the proof of \cref{prop:main counterexample} and its generalisations. 

\begin{prp} 
\label{app:prop:commutator result from spectral gap}
Let $U_1 , \ldots , U_K$ be $K$ unitary matrices on some Hilbert space $\mathcal{H}$ with $d \coloneqq \dim \mathcal{H}$, and suppose that the self-adjoint unital quantum channel $\mathcal{E}$ defined by 
\begin{equation*}
\mathcal{E} (X) \coloneqq \frac{1}{2K} \sum_{j=1}^K \left( U_j X U_j^\dagger + U_j^\dagger X U_j \right) ,
\end{equation*}
has second-largest eigenvalue equal to $\lambda_2$ (which must be real and satisfy $\lambda_2 \leq \lambda_1 = 1$ since $\mathcal{E}$ is self-adjoint and is a quantum channel). Then, for all matrices $X \in \mathcal{B}(\mathcal{H})$, we have
\begin{equation*}
(1-\lambda_2) \left( \| X \|_2^2 - \frac{| \Tr X |^2}{d} \right) \leq \frac{1}{2K} \sum_{j=1}^K \left\| \left[ X , U_j \right] \right\|_2^2 .
\end{equation*}
\end{prp}

\begin{proof}
We note first that the (finite-dimensional) channel $\mathcal{E}$ is self-adjoint w.r.t. the \emph{Frobenius inner product} $\langle X , Y \rangle \coloneqq \Tr X^\dagger Y$, and thus diagonalizable. Secondly, if $Y$ is a traceless matrix, it must be orthogonal to $\id_{\mathcal H}$ since $\Tr \id_{\mathcal H} Y = \Tr Y = 0$, and thus belong to the union of eigenspaces for $\mathcal{E}$ with eigenvalues less than or equal to $\lambda_2$, since $\id_{\mathcal H}$ is the eigenvector associated with the largest eigenvalue $1$ of the unital channel $\mathcal{E}$. Thus, from the maximization principle for finite self-adjoint linear maps and by using the cyclicity of the trace and the fact that the $U_j$'s are unitaries, we get
\begin{equation} 
\label{eq:app:proof commutator s1}
\begin{aligned}
\frac{1}{2K}\sum_{j=1}^K\left\|[Y,U_j]\right\|_2^2
=\|Y\|_2^2-\langle Y,\mathcal{E}(Y)\rangle \geq(1-\lambda_2)\|Y\|_2^2 .
\end{aligned}
\end{equation}
Now, let $X \in \mathcal{B}(\mathcal{H})$ be an arbitrary matrix. Then, clearly $X - \id_{\mathcal H} \Tr (X) / d$ is traceless, and using the result derived in \cref{eq:app:proof commutator s1} gives us
\begin{equation*}
\begin{split}
(1-\lambda_2) \left( \left\| X \right\|_2^2 - \frac{| \Tr X |^2}{d} \right) 
  &= (1-\lambda_2) \left\| X - \id_{\mathcal H} \frac{\Tr X}{d} \right\|_2^2 \\
  &\leq \frac{1}{2K} \sum_{j=1}^K \left\| \left[ X - \id_{\mathcal H} \frac{\Tr X}{d} , U_j \right] \right\|_2^2 = \frac{1}{2K} \sum_{j=1}^K \left\| \left[ X , U_j \right] \right\|_2^2 ,
\end{split}
\end{equation*}
thus proving the proposition.
\end{proof}

\section{Extending \texorpdfstring{\cref{prop:main counterexample}}{prop:main counterexample} to \texorpdfstring{$|A|=2,3$}{A=2,3} } \label{app:extension to d=2 and 3 cases}

We here extend \cref{prop:main counterexample} to the cases $|A|=3$ (\cref{prop:d=3 special case}) and $|A|=2$ (\cref{prop:d=2 case}), using the same random-unitary estimate as in its proof. The qutrit case only changes how the unitaries are placed in the state. For qubits, we take a Hermitian logarithm of one unitary and adjust the witness to account for recovery channels that need not be unital.

\begin{prp}\label{prop:d=3 special case}
Let $|A|=3$, and let $U_1,U_2,U_3\in\mathcal U(B)$ be independent Haar random unitaries. For an orthonormal basis $\{\ket1,\ket2,\ket3\}$ of $A$, define
\begin{equation*}
\rho_{AB}\coloneqq\frac{2\id_{AB}+F}{6|B|},\qquad
F\coloneqq\sum_{j=1}^3\bigl(\ket j\!\bra{j+1}\otimes U_j+\ket{j+1}\!\bra j\otimes U_j^\dagger\bigr),
\end{equation*}
where $\ket4\coloneqq\ket1$. For every $0<\varepsilon<1-\sqrt5/3$, with probability approaching $1$ as $|B|\rightarrow\infty$, every pair of channels $T:\mathcal B(B)\rightarrow\mathcal B(C)$ and $R:\mathcal B(C)\rightarrow\mathcal B(B)$ with $|C|\leq|B|$ satisfies
\begin{equation*}
\left\|\rho_{AB}-\iden_A\otimes(R\circ T)(\rho_{AB})\right\|_1
\geq\frac12\left(1-\frac{\sqrt5}{3}-\varepsilon\right)\left(1-\frac{|C|}{|B|}\right).
\end{equation*}
\end{prp}

\begin{proof}
The block form gives $\Tr_A F=0$, $\Tr_A F^2=6\id_B$ and $\|F\|_\infty\leq2$, so $\rho_{AB}$ is a quantum state and $F/2$ is a valid trace-norm witness. As in the proof of \cref{prop:main counterexample}, choose Kraus operators $K_l$ of $R\circ T$ with rank at most $|C|$. The same calculation gives
\begin{equation*}
\Tr\!\left[\frac F2\bigl(\rho_{AB}-\iden_A\otimes(R\circ T)(\rho_{AB})\bigr)\right]
=\frac1{12|B|}\sum_l\sum_{j=1}^3\|[K_l,U_j]\|_2^2.
\end{equation*}
Hastings' theorem \cite{Hastings2007-ev}, in the form of \cref{app:prop:commutator result from spectral gap}, now with three independent unitaries, and the Kraus-rank estimate from that proof give the stated bound.
\end{proof}

\begin{prp}\label{prop:d=2 case}
Let $|A|=2$, and let $U_1,U_2\in\mathcal U(B)$ be independent Haar random unitaries. Choose a Hermitian logarithm $H$ of $U_1$, so that $U_1=e^{iH}$ and $\|H\|_\infty\leq\pi$, and put $L\coloneqq\pi+1$. In an orthonormal basis of $A$, define
\begin{equation*}
\label{eq:prop:d=2:state 1}
F \coloneqq\begin{pmatrix}H&U_2\\U_2^\dagger&-H\end{pmatrix} 
  = ( \ket{0}\!\bra{0} - \ket{1}\!\bra{1} ) \otimes H + \ket{0}\!\bra{1} \otimes U_2 + \ket{1}\!\bra{0} \otimes U_2^\dagger,
\qquad \rho_{AB}\coloneqq\frac{\id_{AB}+F/L}{2|B|}.
\end{equation*}
Then $\rho_{AB}$ is a quantum state. For every $0<\varepsilon<1-\sqrt3/2$, with probability approaching one as $|B|\rightarrow\infty$, every pair of channels $T:\mathcal B(B)\rightarrow\mathcal B(C)$ and $R:\mathcal B(C)\rightarrow\mathcal B(B)$ with $|C|\leq|B|$ satisfies
\begin{equation}\label{eq:prop:d=2:ineq}
\left\|\rho_{AB}-(\iden_A\otimes(R\circ T))(\rho_{AB})\right\|_1
\geq\frac{2(1-\sqrt3/2-\varepsilon)}{(\pi+1)^2}\left(1-\frac{|C|}{|B|}\right).
\end{equation}
In particular, the prefactor can be chosen arbitrarily close to $2(1-\sqrt3/2)/(\pi+1)^2\approx0.0156213$ in every sufficiently large dimension of $B$.
\end{prp}

\begin{proof}\label{proof:prop:d=2 case}
We have $\Tr_A F=0$ and $\|F\|_\infty\leq\pi+1=L$, which prove positivity and unit trace of $\rho_{AB}$. Use the witness
\begin{equation*}\label{app:eq:qubit corrected witness}
G\coloneqq\frac FL+\id_A\otimes\frac{\pi^2\id_B-H^2}{2L^2}.
\end{equation*}
For $-\pi\leq h\leq\pi$, both scalar functions $\pm h/L+(\pi^2-h^2)/(2L^2)$ lie in $[-\pi/L,\pi/L]$. Thus the diagonal part of $G$ has norm at most $\pi/L$, while its off-diagonal part has norm $1/L$, resulting in $\|G\|_\infty\leq1$. Choose Kraus operators $K_l$ of $R\circ T$ with rank at most $|C|$. Since $\rho_B=\id_B/|B|$ and $\Tr_A F^2=2(H^2+\id_B)$, expansion of the trace gives
\begin{equation}
\label{app:eq:qubit corrected identity}
\begin{split}
 &\left\|\rho_{AB}-(\iden_A\otimes(R\circ T))(\rho_{AB})\right\|_1
\geq\Tr\!\left[G\bigl(\rho_{AB}-(\iden_A\otimes(R\circ T))(\rho_{AB})\bigr)\right]\\
  &\phantom{=====}
   = \frac{1}{2 |B| L^2 } \left[  \Tr \left( F^2 - F \iden_A \otimes (R \circ T) (F) \right)  - \Tr \left( ( F^2 / 2  - 2 \pi^2 \id_{AB} ) ( \id_{AB} - \iden_A \otimes (R \circ T) (\id_{AB}) ) \right) \right] \\
  &\phantom{=====}
   = \frac{1}{4 |B| L^2} \sum_l \Tr \left( F^2 \id_A \otimes  K_l^\dagger K_l  - 2 F ( \id_A \otimes K_l ) F ( \id_A \otimes K_l^\dagger ) + F^2 \id_A \otimes K_l K_l^\dagger \right) \\
   &\phantom{=====}
    =\frac1{4|B|L^2}\sum_l\|[\id_A\otimes K_l,F]\|_2^2.
\end{split}
\end{equation}
The second term in $G$ cancels the contribution involving $(R\circ T)(\id_B)-\id_B$; only trace preservation is used. 

It follows from the triangle inequality and the H\"{o}lder inequality $\| XY \|_2 \leq \| X \|_\infty \| Y \|_2$ that for any $X\in\mathcal B(B)$,
\begin{equation} \label{eq:proof unitary log norm}
[X,U_1]=i\int_0^1e^{itH}[X,H]e^{i(1-t)H}\,dt
\ \  \Rightarrow \ \ \|[X,U_1]\|_2\leq\|[X,H]\|_2.
\end{equation}
Consequently,
\begin{equation*}
\|[\id_A\otimes X,F]\|_2^2
=2\bigl(\|[X,H]\|_2^2+\|[X,U_2]\|_2^2\bigr)
\geq2\sum_{j=1}^2\|[X,U_j]\|_2^2.
\end{equation*}
Hastings' theorem with two independent unitaries gives, simultaneously for all $X$, with probability approaching one,
\begin{equation}
\sum_{j=1}^2\|[X,U_j]\|_2^2
\geq4\left(1-\frac{\sqrt3}{2}-\varepsilon\right)
\left(\|X\|_2^2-\frac{|\Tr X|^2}{|B|}\right).
\end{equation}
Insert this into \cref{app:eq:qubit corrected identity} and use $|\Tr K_l|^2\leq|C|\|K_l\|_2^2$ and $\sum_l\|K_l\|_2^2=|B|$, exactly as in the proof of \cref{prop:main counterexample}. This proves \cref{eq:prop:d=2:ineq}.
\end{proof}

\section{Extending \texorpdfstring{\cref{prop:main counterexample}}{prop:main counterexample} to the case when \texorpdfstring{$A$}{A} is classical} \label{app:extension to classical A}

We here extend \cref{prop:main counterexample} to the case when $\rho_{AB}$ is classical on $A$, which by the arguments from \cref{sec:proof main results} likewise shows that the state $\rho_{AB}$ from \cref{thm:main thm} can be assumed classical on $A$ (when $|A| \geq 3$), by accepting a slightly worse prefactor in the lower bound. Below, \cref{prop:extension classical A 4} deals with the most general case ($|A| \geq 3$), while \cref{prop:extension classical A 8} only holds for $|A| \geq 8$ but gives a better prefactor -- only a $1 / 2$ asymptotic reduction w.r.t. \cref{prop:main counterexample}.

\begin{prp} \label{prop:extension classical A 8}
Let $|A| \geq 8$ and define $S \coloneqq \lfloor |A| / 4 \rfloor$. Let $B$ be an arbitrary additional quantum system and $U_1 , \ldots , U_S \in \mathcal{U} (B)$ be independent Haar random unitaries on $B$. Define finally the quantum state $\rho_{AB}$ as
\begin{equation*}
\begin{aligned}
\rho_{AB} = \frac{1}{ 4 S |B|} \sum_{j=1}^{S} \sum_{a \in \{0,1\} } \left( \vphantom{\frac{1}{1}} \ket{j , a , 0 }\!\bra{j , a , 0 } \otimes \left( \vphantom{2_2^2} \id_B + (-1)^a \Re ( U_j ) \right) + \ket{j , a , 1 }\!\bra{j , a , 1 } \otimes \left( \vphantom{2_2^2} \id_B + (-1)^a \Im ( U_j ) \right) \right) ,
\end{aligned}
\end{equation*}
where $\{ \ket{j,a,b} \}_{a,b \in \{ 0,1 \} , 1 \leq j \leq S }  $ is a set of orthonormal vectors on $A$, and $\Re (U_j) \coloneqq ( U_j + U_j^\dagger ) / 2$ , $\Im (U_j) \coloneqq (U_j - U_j^\dagger) / (2i)$. Then, for every $0<\varepsilon<1-\sqrt{2S-1}/S$, and with probability converging to $1$ as $|B| \rightarrow \infty$, for all pairs of quantum channels $T : \mathcal{B}(B) \rightarrow \mathcal{B}(C)$ and $R : \mathcal{B}(C) \rightarrow \mathcal{B}(B) $ with $|C| \leq |B|$, the following is satisfied
\begin{equation*}
\begin{aligned}
\left\| \vphantom{2_2^2} \rho_{AB} - \iden_A \otimes ( R \circ T) (  \rho_{AB} ) \right\|_1 \geq \frac{1}{2} \left( 1 - \frac{\sqrt{2S-1}}{S} - \varepsilon \right) \left( 1 - \frac{|C|}{|B| } \right)  .
\end{aligned}
\end{equation*} 
\end{prp}

\begin{proof}
It is easy to see that $\rho_{AB}$ is a quantum state, since it follows from the triangle inequality that $\| \Re (U_j) \|_\infty , \| \Im (U_j) \|_\infty \leq 1 $ for any unitary $U_j$. Defining $H \coloneqq 4 S |B| \rho_{AB} - \id_{AB}$, we have $\| H \|_\infty \leq 1$, and using $\id_B = \Re(U_j)^2 + \Im (U_j)^2$ and $U_j X U_j^\dagger Y + U_j^\dagger X U_j Y = 2 (\Re (U_j) X \Re (U_j) Y + \Im (U_j) X \Im (U_j) Y  )$, we get:
\begin{equation} 
\label{eq:proof CQ extension l1}
\begin{split}
\left\| \vphantom{2_2^2} \rho_{AB} - \iden_A \otimes ( R \circ T) (  \rho_{AB} ) \right\|_1  
  &\geq \Tr \left( H (\rho_{AB} - \iden_A \otimes ( R \circ T) (  \rho_{AB} )) \right) \\
  &= \frac{2}{4 S |B|} \sum_{j=1}^S \Tr \left[ \id_B - \sum_l ( \Re (U_j) K_l \Re (U_j) K_l^\dagger + \Im (U_j) K_l \Im (U_j) K_l^\dagger )  \right] \\
  &= \frac{1}{4 S |B|} \sum_l \sum_{j=1}^S \left\| \left[ U_j , K_l \right] \right\|_2^2 ,
\end{split}
\end{equation}
where we have like in the proof of \cref{prop:main counterexample} defined $R \circ T \coloneqq \sum_l K_l \cdot K_l^\dagger $. Like in the proof of \cref{prop:main counterexample}, using Hastings theorem \cite{Hastings2007-ev} in the form of \cref{app:prop:commutator result from spectral gap} in \cref{eq:proof CQ extension l1} above proves the desired result. 
\end{proof}

\begin{prp} \label{prop:extension classical A 4}
Let $|A| \geq 3$ and define $L \coloneqq \lfloor |A| / 3 \rfloor$. Let $B$ be an arbitrary additional quantum system and $U_1 , \ldots , U_{2 L } \in \mathcal{U} (B)$ be independent Haar random unitaries on $B$. Further, let $H_j$ be the Hermitian logarithm of $U_j$, i.e. $U_j = e^{i H_j}$, chosen such that $- \pi \id_B \leq H_j \leq \pi \id_B$. Define the quantum state $\rho_{AB}$ as
\begin{equation*}
\begin{aligned}
\rho_{AB} = \frac{1}{ 4 L |B|} \sum_{j=1}^{L} \left[  \ket{3j-2 } \bra{3j-2 } \otimes \left( \id_B + \frac{ H_{2j-1}}{\pi} \right) + \ket{3j-1 } \bra{3j-1 } \otimes \left( \id_B + \frac{ H_{2j}}{\pi} \right) \right. \\
\left. + \ket{3j } \bra{3j } \otimes \left( 2 \id_B - \frac{ H_{2j-1}+H_{2j}}{\pi} \right) \right]  ,
\end{aligned}
\end{equation*}
where $\{ \ket{j} \}_{1 \leq j \leq 3 L }  $ is a set of orthonormal vectors on $A$. Then, for any $0<\varepsilon<1-\sqrt{4L-1}/(2L)$, we have with probability converging to $1$ as $|B| \rightarrow \infty$, that for all pairs of quantum channels $T : \mathcal{B}(B) \rightarrow \mathcal{B}(C)$ and $R : \mathcal{B}(C) \rightarrow \mathcal{B}(B) $, with $|C| \leq |B|$, the following is satisfied
\begin{equation*} \label{eq:class lower B a=3}
\begin{aligned}
\left\| \vphantom{2_2^2} \rho_{AB} - \iden_A \otimes ( R \circ T) (  \rho_{AB} ) \right\|_1 \geq \frac{2}{7 \pi^2} \left( 1 - \frac{\sqrt{4L-1}}{2L} - \varepsilon \right) \left( 1 - \frac{|C|}{|B| } \right)  .
\end{aligned}
\end{equation*} 
\end{prp}

\begin{proof}
It is easy to see that $\rho_{AB}$ is a quantum state, since all $H_j$ satisfy $\| H_j \|_\infty \leq \pi$. Defining $G$ as
\begin{equation*}
G \coloneqq \frac{1}{\pi} \sum_{j=1}^{L} \left[  \ket{3j-2 } \bra{3j-2 } \otimes  H_{2j-1}  + \ket{3j-1 } \bra{3j-1 } \otimes  H_{2j} \vphantom{2_2^2} \right] ,
\end{equation*}
we have $\| G \|_\infty \leq 1$. Using the relations $\| [X , U_j ] \|_2 = \| [X , e^{i H_j}] \|_2 \leq \| [ X , H_j ] \|_2$ proved in \cref{eq:proof unitary log norm}, \cref{app:extension to d=2 and 3 cases}, we see that
\begin{equation} 
\label{eq:proof CQ extension 3 l1}
\begin{split}
 &\left\| \vphantom{2_2^2} \rho_{AB} - \iden_A \otimes ( R \circ T) (  \rho_{AB} ) \right\|_1 
  \geq \Tr \left[ G (\rho_{AB} - \iden_A \otimes ( R \circ T) (  \rho_{AB} )) \right] \\
  &\phantom{====}
   = \frac{1}{2 \pi^2 |B|} \frac{1}{4 L} \sum_{j=1}^{2L} \left( \sum_l \left( \left\| \left[ H_j , K_l \right] \right\|_2^2 + \Tr \left[ H_j^2 (K_l^\dagger K_l - K_l K_l^\dagger) \right] \right) + 2 \pi \Tr H_j (\id_B - (R \circ T)(\id_B)) \right) \\
  &\phantom{====}
   \geq \frac{1}{2 \pi^2 |B|} \frac{1}{4 L} \sum_l \sum_{j=1}^{2L} \left\| \left[ U_j , K_l \right] \right\|_2^2 - \frac{3}{4} \Delta, \text{ with } \Delta \coloneqq \left\| \id_B - (R \circ T) (\id_B) \right\|_1 / |B| ,
\end{split}
\end{equation}
where (like in the proof of \cref{prop:main counterexample}) we have defined $R \circ T \coloneqq \sum_l K_l \cdot K_l^\dagger $ with $\rnk (K_l) \leq |C|$. Just as in the proof of \cref{prop:main counterexample}, using Hastings' theorem in the final line of \cref{eq:proof CQ extension 3 l1}, while also using that by the monotonicity of the trace norm under quantum channels (specifically $\Tr_A $), we have $\| \rho_{AB} - \iden_A \otimes (R \circ T) (\rho_{AB}) \|_1 \geq \| \rho_B - (R \circ T) (\rho_B) \|_1 = \| \id_B - (R \circ T)(\id_B) \|_1 / |B| \coloneqq \Delta $, we get with probability approaching $1$ as $|B| \rightarrow \infty$ that
\begin{equation*}
\begin{split}
\left\| \vphantom{2_2^2} \rho_{AB} - \iden_A \otimes ( R \circ T) (  \rho_{AB} ) \right\|_1
  &\geq \max \left( \frac{1}{2 \pi^2} \left( 1 - \frac{\sqrt{4L-1}}{2L} - \varepsilon \right) \left( 1 - \frac{|C|}{|B| } \right) - \frac{3}{4} \Delta , \ \Delta \right) \\ 
  &\geq \frac{2}{7 \pi^2} \left( 1 - \frac{\sqrt{4L-1}}{2L} - \varepsilon \right) \left( 1 - \frac{|C|}{|B| } \right) .
\end{split}
\end{equation*} 
\end{proof}

\section{A stronger random-state counterexample}
\label{app:random states}

We give a random-state construction whose loss of mutual information has an asymptotically optimal leading coefficient. The output may contain an arbitrary finite classical record in addition to a small quantum system. This thus constitutes a strengthening and generalization of \cref{cor:tensor prod construction mutual info}, but with a much more complicated proof. 

\begin{thm}\label{app:thm:random states}
For every $0<\beta<1$, there is an integer $s_\beta>1$, which can be chosen to satisfy
\begin{equation}\label{app:eq:random s growth}
s_\beta\leq C_0\beta^{-2}\frac{\log(C_0/\beta)}{\log e}
\end{equation}
for a universal constant $C_0$, with the following property. Fix $|A|\geq\max\{8,\sqrt{2s_\beta}\}$, let $|E|=s_\beta|B|/|A|$ and put $d=\lfloor |E|/|A|\rfloor$. Draw a unit vector $\ket{\psi}_{ABE}$ according to Haar measure and set $\rho_{AB}=\Tr_E\ket{\psi}\bra{\psi}$. With probability tending to one as $|B|\rightarrow\infty$, along dimensions for which $|E|$ is an integer, we have simultaneously
\begin{align}
I(A:B)_\rho&\geq2\log|A|-\log s_\beta-2,\label{app:eq:random input}\\
I(A:XC)_{(\iden_A\otimes T)(\rho)}&\leq\beta\log|A|+5\log\log|A|+13\label{app:eq:random output}
\end{align}
for every quantum channel $T:\mathcal{B}(B)\rightarrow\mathcal{B}(XC)$ whose output $X$ is classical and whose conditional quantum output has dimension at most $|C|\leq d$. There is no bound on the finite cardinality of $X$.
\end{thm}

The first step is to prove a lower bound on the conditional entropy after every projection of rank at most $d$. We then pass from projections to arbitrary Kraus operators by spectral pinching. For a nonzero projector $P$ on $B$, write
\begin{equation*}\label{app:eq:normalized projection}
\rho^P_{AB}\coloneqq\frac{(\id_A\otimes P)\rho_{AB}(\id_A\otimes P)}{\Tr(P\rho_B)}.
\end{equation*}
We use the conditional-entropy continuity bound from \cite{Winter2016_continuitybounds}: for states $\omega_{AC},\eta_{AC}$ with $\|\omega-\eta\|_1/2\leq t\leq1$,
\begin{equation}\label{app:eq:conditional continuity}
\left|H(A|C)_\omega-H(A|C)_\eta\right|
\leq2t\log|A|+g(t),
\end{equation}
where $g$ is defined in \cref{lemma:hypothesis are equivalent}, with $0\log0=0$. 

\subsection{Uniform bounds for projected states}

\begin{lem}\label{app:lem:conditional Lipschitz}
For unit vectors $\ket\psi,\ket\varphi$ on $ACE$, with $|A|\geq2$,
\begin{equation*}\label{app:eq:conditional Lipschitz}
\left|H(A|C)_\psi-H(A|C)_\varphi\right|
\leq\frac{\pi}{\sqrt2}(1+\log|A|)\,\|\ket\psi-\ket\varphi\|_2.
\end{equation*}
\end{lem}
\begin{proof}
The fidelity continuity bound of \cite[Theorem 1]{Tan2026-fc}, converted to base two, and monotonicity of fidelity under partial trace give
\begin{equation*}
\left|H(A|C)_\psi-H(A|C)_\varphi\right|
\leq2\theta\log\left(|A|+\sqrt{|A|^2-1}\right),
\qquad\theta=\arccos|\braket{\psi|\varphi}|.
\end{equation*}
Aligning the phase of $\ket\varphi$ with $\ket\psi$ cannot increase their Euclidean distance. Since $0\leq\theta\leq\pi/2$, the arc-to-chord bound gives $\theta\leq\pi\|\ket\psi-\ket\varphi\|_2/(2\sqrt2)$. Finally, $|A|+\sqrt{|A|^2-1}\leq2|A|$.
\end{proof}

\begin{lem}\label{app:lem:fixed projection}
Fix $0<\beta<1$ and a rank-$r$ projector $P$, with $|A|r\leq|E|$. For $\alpha_0=\frac{1}{9\pi^5}$,
\begin{equation}\label{app:eq:fixed projection probability}
\begin{aligned}
&\Pr\left\{H(A|B)_{\rho^P}<\left(1-\frac\beta2\right)\log|A|-\frac{|A|r\log e}{2|E|}\right\}\\
&\hspace{15mm}\leq2\exp\left[-\alpha_0\beta^2|A|r|E|\left(\frac{\log|A|}{1+\log|A|}\right)^2\right].
\end{aligned}
\end{equation}
\end{lem}
\begin{proof}
The normalized projected vector is Haar distributed on $A\otimes\operatorname{ran}P\otimes E$. Since $H(B)_{\rho^P}\leq\log r$, the mean entropy bound in \cite[Lemma II.4]{Hayden2006-ns} gives
\begin{equation*}
\mathbb E H(A|B)_{\rho^P}\geq\mathbb E H(AB)_{\rho^P}-\log r
\geq\log|A|- \frac{|A|r\log e}{2|E|}.
\end{equation*}
Apply the mean form of Levy's lemma \cite[Lemma III.1]{Hayden2006-ns} on the real sphere of dimension $2|A|r|E|-1$, with the Lipschitz constant from \cref{app:lem:conditional Lipschitz} and deviation $(\beta/2)\log|A|$. This gives \cref{app:eq:fixed projection probability} with the stated universal constant.
\end{proof}

\begin{lem}\label{app:lem:projection net}
For $0<\zeta<1$, the rank-$r$ projectors on $B$ have an operator-norm $\zeta$-net of cardinality at most $(5/\zeta)^{2|B|r}$.
\end{lem}
\begin{proof}
Regard the $|B|\times r$ isometries as a subset of the operator-norm unit ball in a real vector space of dimension $2|B|r$. A maximal $\zeta/2$-separated subset is a $\zeta/2$-net and has size at most $(1+4/\zeta)^{2|B|r}$, by comparing volumes of disjoint balls of radius $\zeta/4$, see also~\cite[Lemma 5.8]{AliceBobBanach}. 
For isometries $\Gamma,\Gamma'$, we have $\|\Gamma\Gamma^\dagger-\Gamma'\Gamma'^\dagger\|_\infty\leq2\|\Gamma-\Gamma'\|_\infty$. Passing to range projections and using $1+4/\zeta\leq5/\zeta$ proves the claim.
\end{proof}

\begin{lem}\label{app:lem:flat marginal}
For the Haar construction in \cref{app:thm:random states},
\begin{equation*}\label{app:eq:flat marginal probability}
\Pr\left\{\frac{\id_B}{2|B|}\leq\rho_B\leq\frac{3\id_B}{2|B|}\right\}
\geq1-2\cdot10^{2|B|}e^{-|A||E|\log e/96}.
\end{equation*}
\end{lem}
\begin{proof}
For a fixed unit vector $\ket v$ on $B$, apply the projector-overlap estimate \cite[Lemma III.5]{Hayden2006-ns} to $\ket\psi\bra\psi$ and the rank-$|A||E|$ projector $\id_A\otimes\ket v\bra v\otimes \id_E$. Its two tails, with relative deviation $1/4$, give
\begin{equation*}
\Pr\left\{\big||B|\bra v\rho_B\ket v-1\right|>\frac14\big\}
\leq 2e^{-|A||E|\log e/96}.
\end{equation*}
Set $\Delta=\rho_B-\id_B/|B|$. Take a trace-norm $1/2$-net of pure states of size at most $10^{2|B|}$ \cite[Lemma III.6]{Hayden2006-ns}. Except on an event of probability at most $2\cdot10^{2|B|}e^{-|A||E|\log e/96}$, every net state $\tau$ satisfies $|\Tr\Delta\tau|\leq1/(4|B|)$. The operator norm of the Hermitian $\Delta$ is achieved on a pure state $\sigma$. Choosing a net state $\tau$ with $\|\sigma-\tau\|_1\leq1/2$ yields
\begin{equation*}
\|\Delta\|_\infty\leq|\Tr\Delta\tau|+\|\sigma-\tau\|_1\|\Delta\|_\infty
\leq\frac1{4|B|}+\frac12\|\Delta\|_\infty.
\end{equation*}
Rearranging gives the claim.
\end{proof}

\begin{lem}\label{app:lem:all projections}
The integer $s_\beta$ in \cref{app:thm:random states} can be chosen so that, with probability tending to one as $|B|\rightarrow\infty$,
\begin{align}
\frac{\id_B}{2|B|}&\leq\rho_B  \leq\frac{3\id_B}{2|B|},\label{app:eq:random flatness}\\
H(A|B)_{\rho^P}&\geq(1-\beta)\log|A|-1
\quad\text{for all }1\leq\rnk P\leq d.\label{app:eq:all projection entropy}
\end{align}
\end{lem}
\begin{proof}
First choose the nets from \cref{app:lem:projection net}, for each $1\leq r\leq d$, independently of the state. Since $|A|r\leq|E|$, \cref{app:lem:fixed projection} and the union bound show that the probability that any net projector violates
\begin{equation*}
H(A|B)_{\rho^P}\geq(1-\beta/2)\log|A|-\frac{\log e}{2}
\end{equation*}
is at most
\begin{equation}\label{app:eq:projection net failure}
2\sum_{r=1}^{d}\exp\left[-|B|r\left(\alpha_0\beta^2s_\beta\left(\frac{\log|A|}{1+\log|A|}\right)^2-\frac{2\log(5/\zeta)}{\log e}\right)\right].
\end{equation}
To transfer this to all projectors, let $\ket\psi$ purify $\rho$ and put $\ket x=(\id_A\otimes P\otimes \id_E)\ket\psi$, $\ket y=(\id_A\otimes Q\otimes \id_E)\ket\psi$, $p=\|\ket x\|_2^2$ and $q=\|\ket y\|_2^2$. On the event in \cref{app:eq:random flatness}, projectors $P,Q$ of the same rank $r$ with $\|P-Q\|_\infty\leq\zeta$ satisfy
\begin{equation*}
\|\ket x-\ket y\|_2^2\leq\frac{3r}{|B|}\zeta^2,
\qquad p,q\geq\frac r{2|B|}.
\end{equation*}
Consequently,
\begin{equation*}
\left\|\frac{\ket x}{\sqrt p}-\frac{\ket y}{\sqrt q}\right\|_2^2
=\frac{\|\ket x-\ket y\|_2^2-(\sqrt p-\sqrt q)^2}{\sqrt{pq}}
\leq6\zeta^2.
\end{equation*}
Choose $\zeta=\beta/40$. Since $1+\log|A|\leq(5/3)\log|A|$, \cref{app:lem:conditional Lipschitz} bounds the conditional-entropy change by $(\beta/4)\log|A|$. Choose $s_\beta$ to be the smallest integer satisfying
\begin{equation}\label{app:eq:choose random s}
s_\beta\geq\frac{96(2\log10+\log e)}{(\log e)^2},
\qquad\frac9{25}\alpha_0\beta^2s_\beta\geq\frac{2\log(200/\beta)}{\log e}+1.
\end{equation}
The failure probability in \cref{app:eq:projection net failure} is then at most $2e^{-|B|}/(1-e^{-|B|})$, and \cref{app:lem:flat marginal} gives a further failure probability at most $2e^{-|B|}$ for \cref{app:eq:random flatness}. Transfer from the nets gives
\begin{equation*}
H(A|B)_{\rho^P}\geq(1-3\beta/4)\log|A|-\frac{\log e}{2}
\geq(1-\beta)\log|A|-1,
\end{equation*}
establishing \cref{app:eq:all projection entropy}. The least integer satisfying \cref{app:eq:choose random s} also obeys \cref{app:eq:random s growth} for a universal $C_0$.
\end{proof}

\subsection{From projections to arbitrary channels}

\begin{lem}\label{app:lem:coarse pinching}
Let $\omega_{AC}$ be a state and let $\tau_C$ be a state with the same support as $\omega_C$, with $\kappa^{-1}\tau_C\leq\omega_C\leq\kappa\tau_C$ for some $\kappa\geq1$. For $b>1$, let $\Pi_j$ be the spectral projections of $\tau_C$ for $(b^{-(j+1)},b^{-j}]$, $j\geq0$, and put
\begin{equation*}
M=|A|^2\kappa^2b,\qquad
G=\sum_{\ell\geq1}\ell\min\{1,M b^{-\ell/2}\}.
\end{equation*}
For every positive integer $m$, consecutive bins can be collected into groups of at most $m$ bins so that the corresponding pinching $\mathcal P$ satisfies
\begin{equation}\label{app:eq:coarse distance}
\frac12\|\omega-(\iden_A\otimes\mathcal P)(\omega)\|_1\leq G/m.
\end{equation}
If $\mathcal F$ pinches into the individual bins, then, whenever $G/m\leq\epsilon\leq1$,
\begin{equation}\label{app:eq:pinching information}
I(A:C)_\omega\leq I(A:C)_{(\iden_A\otimes\mathcal F)(\omega)}
+2\epsilon\log|A|+g(\epsilon)+\log m.
\end{equation}
Moreover, with $J_0=\lceil\log M/(-\log b^{-1/2})\rceil$,
\begin{equation}\label{app:eq:series bound}
G\leq\frac{J_0(J_0+1)}2+\frac{J_0b^{-1/2}}{1-b^{-1/2}}+\frac{b^{-1/2}}{(1-b^{-1/2})^2}.
\end{equation}
\end{lem}
\begin{proof}
Restrict $C$ to the common support. Pinching in a basis of $A$ gives $\omega\leq|A|\sum_i\ket i\!\bra i\otimes\omega_{ii}\leq|A|\id_A\otimes\omega_C\leq|A|\kappa \id_A\otimes\tau_C$. Write $\omega_{j\ell}=(\id_A\otimes\Pi_j)\omega(\id_A\otimes\Pi_\ell)$ and $p_j=\Tr\omega_{jj}$. Positivity and the (Schatten) Cauchy--Schwarz inequality yield
\begin{equation*}
B_\ell\coloneqq\sum_j\|\omega_{j,j+\ell}\|_1
\leq\sum_j\sqrt{p_jp_{j+\ell}}\leq1.
\end{equation*}
Indeed, with $R_j=\id_A\otimes\Pi_j$, the off-diagonal block equals $(R_j\sqrt\omega)(R_{j+\ell}\sqrt\omega)^\dagger$, and $\|XY^\dagger\|_1\leq\|X\|_2\|Y\|_2$.

There is also a decaying bound. Write $\omega=|A|\kappa(\id_A\otimes\sqrt\tau)Z(\id_A\otimes\sqrt\tau)$, where $0\leq Z\leq \id_{AC}$. Then
\begin{equation*}
\|\omega_{j,j+\ell}\|_\infty\leq|A|\kappa b^{-j-\ell/2},
\qquad\rnk\omega_{j,j+\ell}\leq|A|\rnk\Pi_j.
\end{equation*}
The lower endpoint of a bin and $\tau_C\leq\kappa\omega_C$ imply $(\rnk\Pi_j)b^{-j}\leq b\Tr(\tau_C\Pi_j)\leq b\kappa p_j$. Hence $\|\omega_{j,j+\ell}\|_1\leq Mb^{-\ell/2} p_j$, giving $B_\ell\leq\min\{1,Mb^{-\ell/2}\}$. For $\ell>J_0$, $Mb^{-\ell/2}\leq b^{-(\ell-J_0)/2}$; summing this arithmetic--geometric tail proves \cref{app:eq:series bound}.

For $h\in\{0,\ldots,m-1\}$, group bin $j$ by $\lfloor(j+h)/m\rfloor$. Bins separated by $\ell$ lie in different groups for a fraction $\min\{\ell/m,1\}$ of these shifts. Counting both off-diagonal halves and the factor $1/2$ in trace distance gives
\begin{equation*}
\frac1m\sum_{h=0}^{m-1}\frac12\|\omega-(\iden_A\otimes\mathcal P_h)(\omega)\|_1
\leq\sum_{\ell\geq1}\min\{\ell/m,1\}B_\ell\leq G/m.
\end{equation*}
One shift therefore satisfies \cref{app:eq:coarse distance}. There are finitely many nonempty bins; adding empty bins makes this count valid also at the first and last groups.

For that shift, set $V_h=\sum_j e^{2\pi\mathrm i(j+h)/m}\Pi_j$ on the support. The phases are distinct within each coarse group, so
\begin{equation*}
\mathcal F(\mathcal P_h(\omega))=\mathcal F(\omega)
=\frac1m\sum_{\ell=0}^{m-1}(\id_A\otimes V_h^\ell)\mathcal P_h(\omega)(\id_A\otimes V_h^{-\ell}),
\end{equation*}
where the local identity maps are left implicit in the pinching notation. Thus refinement increases $H(AC)$ by at most $\log m$ and cannot decrease $H(C)$. The coarse pinching preserves $\omega_A$, so \cref{app:eq:conditional continuity} bounds its mutual-information loss. This proves \cref{app:eq:pinching information}. Neither the number of bins nor the smallest nonzero eigenvalue enters the bound.
\end{proof}

\begin{lem}\label{app:lem:flat filter}
For a state $\eta_{AC}$ and an operator $\id_C\leq F\leq\Lambda \id_C$, with $\Lambda\geq1$, its normalized filtered state
\begin{equation*}
\eta^F_{AC}=\frac{(\id_A\otimes F)\eta_{AC}(\id_A\otimes F)}{\Tr(F^2\eta_C)}
\end{equation*}
satisfies $\|\eta-\eta^F\|_1/2\leq(\Lambda-1)/(\Lambda+1)$.
\end{lem}
\begin{proof}
Apply $F$ to a purification of $\eta$, and put $\mu=\langle F\rangle$, $\nu=\langle F^2\rangle$. Since $F^2+\Lambda \id_C\leq(\Lambda+1)F$, we have $2\sqrt{\Lambda\nu}\leq\nu+\Lambda\leq(\Lambda+1)\mu$. The overlap of the original and normalized filtered purifications is therefore at least $2\sqrt\Lambda/(\Lambda+1)$. Their trace distance is at most $(\Lambda-1)/(\Lambda+1)$, and partial trace cannot increase it.
\end{proof}

\begin{prp}\label{app:prp:general channels}
Suppose $|A|\geq8$, $\rho_{AB}$ satisfies \cref{app:eq:random flatness}, and $H(A|B)_{\rho^P}\geq h$ for every projector of rank between $1$ and $d$. Then every channel as in \cref{app:thm:random states} satisfies
\begin{equation*}\label{app:eq:general channel bound}
I(A:XC)_{(\iden_A\otimes T)(\rho)}\leq H(A)_\rho-h+5\log\log|A|+12.
\end{equation*}
\end{prp}
\begin{proof}
Refine the output by retaining each Kraus label. Discarding that extra classical label recovers the original output and cannot increase mutual information. Embedding all conditional quantum outputs into a common $d$-dimensional space, we may therefore consider
\begin{equation*}
\sigma_{AXC}=\sum_x\ket x\bra x\otimes(\id_A\otimes K_x)\rho_{AB}(\id_A\otimes K_x^\dagger).
\end{equation*}
For each nonzero branch, put
\begin{align*}
p_x&=\Tr(K_x\rho_BK_x^\dagger),\qquad
\tau^x_C=\frac{K_xK_x^\dagger}{\Tr K_xK_x^\dagger},\\
\omega^x_{AC}&=\frac{(\id_A\otimes K_x)\rho_{AB}(\id_A\otimes K_x^\dagger)}{p_x}.
\end{align*}
The marginal bound \cref{app:eq:random flatness} implies $\tau^x_C/3\leq\omega^x_C\leq3\tau^x_C$. In particular, for $\tau_{XC}=\sum_xp_x\ket x\bra x\otimes\tau^x_C$, we have $\tau_{XC}/3\leq\sigma_{XC}\leq3\tau_{XC}$, including arbitrarily unlikely branches.

Fix $\Lambda>1$ and $0<\epsilon\leq1$. Apply \cref{app:lem:coarse pinching} to $\sigma_{AXC}$, treating $XC$ as its second system, with $\kappa=3$, $b=\Lambda^2$ and an integer $m\geq G/\epsilon$. Its spectral projections preserve $X$. Let $\widehat\sigma$ be the finely pinched state with the bin label $J$ retained. Copying that label is an isometry on the block-diagonal state, so
\begin{equation}\label{app:eq:coarse fine bound}
I(A:XC)_\sigma\leq I(A:XJC)_{\widehat\sigma}
+2\epsilon\log|A|+g(\epsilon)+\log m.
\end{equation}
Within branch $x$, the fine-bin projector $\Pi_j^x$ is a spectral projector of $p_x\tau_C^x$. Thus all nonzero singular values of $\Pi_j^xK_x$ have ratio at most $\Lambda$. Its polar decomposition is an isometry after a positive filter supported on the input projector $P_{xj}=\operatorname{supp}(K_x^\dagger\Pi_j^xK_x)$, of rank at most $d$. After rescaling, the filter has spectrum in $[1,\Lambda]$ and acts on $\rho^{P_{xj}}$. By \cref{app:lem:flat filter,app:eq:conditional continuity},
\begin{equation*}
H(A|C)_{\widehat\sigma^{xj}}\geq h-2t\log|A|-g(t),
\qquad t=\frac{\Lambda-1}{\Lambda+1}.
\end{equation*}
The postselected $A$ marginals need not coincide. However, the complete instrument preserves $\rho_A$, since $\sum_{x,j}K_x^\dagger\Pi_j^xK_x=\id_B$. Averaging the conditional entropies and using \cref{app:eq:coarse fine bound} yields
\begin{equation}\label{app:eq:channel parameters}
I(A:XC)_{(\iden_A\otimes T)(\rho)}\leq H(A)_\rho-h
+2(t+\epsilon)\log|A|+g(t)+g(\epsilon)+\log m.
\end{equation}

Finally set $z=\log|A|\geq3$ and choose $\Lambda=1+1/z$, $\epsilon=1/z$, $t=1/(2z+1)$ and $m=\lceil80z^5\rceil$. Then $q=z/(z+1)$, $M=9|A|^2\Lambda^2\leq36|A|^2$, $\log M\leq5z\log e$ and $-\log q\geq(\log e)/(2z)$. Hence $J_0\leq11z^2$ in \cref{app:eq:series bound} and
\begin{equation*}
G\leq\frac{J_0(J_0+1)}2+J_0z+z(z+1)\leq79z^4.
\end{equation*}
Therefore $G/m\leq\epsilon$. Also $2(t+\epsilon)z\leq3$, and monotonicity of $g$ gives $g(t)+g(\epsilon)\leq g(1/7)+g(1/3)<2$. Since $\log m\leq\log81+5\log z<7+5\log z$, \cref{app:eq:channel parameters} proves the bound. The constants do not depend on $|B|$, $d$, the number of Kraus operators or their smallest nonzero singular values.
\end{proof}

\begin{proof}[Proof of \cref{app:thm:random states}]
Choose $s_\beta$ as in \cref{app:lem:all projections}. On its high-probability event, apply \cref{app:prp:general channels} with $h=(1-\beta)\log|A|-1$ and use $H(A)_\rho\leq\log|A|$. This gives \cref{app:eq:random output}, simultaneously for every channel, without a union bound over channels.

Applying \cref{app:lem:flat marginal} also with $A$ and $B$ exchanged shows that, except on events of total probability at most
\begin{equation*}
2\cdot10^{2|B|}e^{-|A||E|\log e/96}
+2\cdot10^{2|A|}e^{-|B||E|\log e/96},
\end{equation*}
we have $\rho_B\leq3\id_B/(2|B|)$ and $\rho_A\leq3\id_A/(2|A|)$. Purity and $H(E)_\psi\leq\log|E|$ then give
\begin{align*}
I(A:B)_\rho&=H(A)_\rho+H(B)_\rho-H(E)_\psi\\
&\geq\log|A|+\log|B|-\log|E|-2\log(3/2)\\
&\geq2\log|A|-\log s_\beta-2.
\end{align*}
These failure probabilities tend to zero by the choice of $s_\beta$. Intersecting the events proves \cref{app:eq:random input} together with the channel bound.
\end{proof}

\begin{cor}\label{app:cor:near optimal random}
For every sufficiently large fixed $A$, there is an integer $s(|A|)$ with
\begin{equation*}
s(|A|)\leq C_0(\log|A|)^2\frac{\log(C_0\log|A|)}{\log e}
\end{equation*}
such that the Haar construction with $|E|=s(|A|)|B|/|A|$ satisfies, with probability tending to one as $|B|\rightarrow\infty$,
\begin{align*}
I(A:B)_\rho&\geq2\log|A|-3\log\log|A|-C_1,\\
I(A:XC)_{(\iden_A\otimes T)(\rho)}&\leq5\log\log|A|+14
\end{align*}
for every channel with arbitrary finite classical output $X$ and conditional quantum output dimension at most $\lfloor s(|A|)|B|/|A|^2\rfloor$. Here $C_0,C_1$ are universal constants. Thus the information loss is at least $2\log|A|-8\log\log|A|-C_1-14$. In particular, for each fixed target dimension $M$, choosing $B$ sufficiently large gives this loss for all channels with $|C|\leq M$.
\end{cor}
\begin{proof}
Take $\beta=1/\log|A|$ in \cref{app:thm:random states}. Its bound on $s_\beta$ gives the stated growth of $s(|A|)$, and $|A|^2\geq2s(|A|)$ for all sufficiently large $|A|$. Moreover $\log s(|A|)\leq3\log\log|A|+O(1)$. Substituting into \cref{app:eq:random input,app:eq:random output} proves the two estimates. The limit in $|B|$ is taken after fixing $A$, so the output dimension bound eventually exceeds any fixed $M$.
\end{proof}

The same construction gives a classical-reference variant. Let $\mathcal D_A$ dephase $A$ in any fixed basis and put $\widetilde\rho_{AB}=(\mathcal D_A\otimes\iden_B)(\rho_{AB})$. Pinching increases $H(AB)$ by at most $\log|A|$ and cannot decrease $H(A)$, so $I(A:B)_{\widetilde\rho}\geq I(A:B)_\rho-\log|A|$. Data processing also gives $I(A:XC)_{(\iden_A\otimes T)(\widetilde\rho)}\leq I(A:XC)_{(\iden_A\otimes T)(\rho)}$. Hence a state classical on $A$ already forces a loss of at least $\log|A|-8\log\log|A|-C_1-14$ under the same memory constraint.

\section{The information bottleneck and squashed entanglement}
\label{app:bottleneck squashed}

We relate a binary quantum information bottleneck problem to squashed entanglement. The system being compressed is again $B$, and its compressed output is $C$. For a state $\rho_{AB}$ with purification $\ket\psi_{ABE}$, define the bottleneck function \cite{Datta2019-bq} by
\begin{equation}\label{app:eq:QIB definition}
\mathsf I_\rho(R)\coloneqq
\sup_{\substack{T:\mathcal B(B)\rightarrow\mathcal B(C)\\ I(AE:C)_\omega\leq R}}
I(A:C)_\omega,
\qquad\omega_{AEC}=(\iden_{AE}\otimes T)(\ket\psi\!\bra\psi),
\end{equation}
where the supremum allows arbitrary finite-dimensional $C$. Write $\mathsf I_{\rho,D}(R)$ for the same optimization with $|C|\leq D$. The cost is $I(AE:C)$, including the purifying system $E$, rather than only $I(A:C)$.

For a state $\sigma_{AE}$, let
\begin{equation*}\label{app:eq:bounded squashed definition}
E_D(\sigma_{AE})\coloneqq
\min_{\substack{\Tr_C\omega_{AEC}=\sigma_{AE}\\ |C|\leq D}}
\frac12 I(A:E|C)_\omega,
\qquad E_{\mathrm{sq}}(\sigma_{AE})=\inf_{D\geq1}E_D(\sigma_{AE}).
\end{equation*}
The minimum at each finite $D$ exists by compactness and continuity. This is the dimension-restricted squashed entanglement used in \cite{ahmed2026squashedentanglement}.

Let $\ket{u_0},\ket{u_1}$ be pure qubit states with real overlap $\braket{u_0|u_1}=c$, where $0<c<1$, and take the binary source
\begin{equation}\label{app:eq:binary source}
\rho_{AB}=\frac12\ket0\!\bra0_A\otimes\ket{u_0}\!\bra{u_0}_B
+\frac12\ket1\!\bra1_A\otimes\ket{u_1}\!\bra{u_1}_B.
\end{equation}
Here $A$ is classical, $B$ is a qubit, and $H(B)_\rho=h((1+c)/2)$, where $h$ is the binary entropy. Define also the dephased Bell state on $AE$,
\begin{equation}\label{app:eq:dephased Bell}
\sigma^{(c)}_{AE}=\frac12\left(\ket{00}\bra{00}+\ket{11}\bra{11}
+c\ket{00}\bra{11}+c\ket{11}\bra{00}\right).
\end{equation}

\begin{prp}\label{app:prp:bottleneck identity}
For the source in \cref{app:eq:binary source},
\begin{equation}\label{app:eq:bottleneck identity}
\mathsf I_\rho(H(B)_\rho)=1-E_{\mathrm{sq}}(\sigma^{(c)}_{AE}).
\end{equation}
More precisely, for every positive integer $D$,
\begin{equation}\label{app:eq:finite bottleneck comparison}
\mathsf I_{\rho,D}(H(B)_\rho)
\leq1-E_D(\sigma^{(c)}_{AE})
\leq\mathsf I_{\rho,3D}(H(B)_\rho).
\end{equation}
\end{prp}
\begin{proof}
A purification of \cref{app:eq:binary source} is
\begin{equation*}
\ket\psi_{ABE}=\frac1{\sqrt2}\left(\ket0_A\ket{u_0}_B\ket0_E
+\ket1_A\ket{u_1}_B\ket1_E\right).
\end{equation*}
Its $AE$ marginal is $\sigma^{(c)}_{AE}$, so every extension $\omega_{AEC}$ of this state is generated by a channel on $B$. Positivity forces every such extension to be supported on $\operatorname{span}\{\ket{00},\ket{11}\}_{AE}\otimes C$. Consequently $\omega_{AC}$ and $\omega_{EC}$ are identical up to relabelling. Using $H(A)=H(E)=1$ and $H(AE)=H(B)_\rho$, expansion of the entropies gives
\begin{equation}\label{app:eq:bottleneck entropy identity}
I(A:E|C)_\omega
=2-H(B)_\rho+I(AE:C)_\omega-2I(A:C)_\omega.
\end{equation}
If the bottleneck cost satisfies $I(AE:C)_\omega\leq H(B)_\rho$, then $I(A:C)_\omega\leq1-I(A:E|C)_\omega/2$. For $|C|\leq D$, this is at most $1-E_D(\sigma^{(c)}_{AE})$, proving the first inequality in \cref{app:eq:finite bottleneck comparison}. Without a dimension restriction it also proves the upper bound in \cref{app:eq:bottleneck identity}.

Conversely, take any extension $\omega_{AEC}$ with $|C|\leq D$ and set $q=I(A:E|C)_\omega$. As $\rnk\omega_{AEC}\leq2D$, it has a purification on $AECF$ with $|F|\leq2D$. This purification is obtained by an isometry from $B$ to $CF$. Keeping $C$ or $F$ gives complementary channels $T$ and $T^c$. Purity and the equality of the $AC$ and $EC$ marginals give
\begin{align*}
I(AE:C)_\omega+I(AE:F)_\omega&=2H(B)_\rho,\\
I(A:C)_\omega+I(A:F)_\omega&=2-q.
\end{align*}
Choose the two channels with equal probabilities and retain the choice in the direct-sum output $C'=C\oplus F$:
\begin{equation*}
\widetilde T(X)=\frac12T(X)\oplus\frac12T^c(X).
\end{equation*}
The flag is independent of $AE$. Therefore the cost of $\widetilde T$ is exactly $H(B)_\rho$ and its relevance is $1-q/2$, while $|C'|\leq3D$. Choosing an extension attaining $E_D$ proves the second inequality in \cref{app:eq:finite bottleneck comparison}. Finally, taking extensions approaching the squashed-entanglement infimum proves \cref{app:eq:bottleneck identity}.
\end{proof}

\begin{cor}\label{app:cor:binary nonattainment}
For the binary qubit source in \cref{app:eq:binary source}, no finite-dimensional compressed output attains $\mathsf I_\rho(H(B)_\rho)$. For every positive integer $D$,
\begin{equation*}
\mathsf I_{\rho,D}(H(B)_\rho)
<\mathsf I_{\rho,9D+6}(H(B)_\rho)
\leq\mathsf I_\rho(H(B)_\rho).
\end{equation*}
\end{cor}
\begin{proof}
Ahmed and Smith prove that $E_{3D+2}(\sigma^{(c)}_{AE})<E_D(\sigma^{(c)}_{AE})$ for every $D$ \cite[Theorem 2]{ahmed2026squashedentanglement}. Thus \cref{app:eq:finite bottleneck comparison} gives
\begin{equation*}
\begin{aligned}
\mathsf I_{\rho,D}(H(B)_\rho)
&\leq1-E_D(\sigma^{(c)}_{AE})\\
&<1-E_{3D+2}(\sigma^{(c)}_{AE})
\leq\mathsf I_{\rho,9D+6}(H(B)_\rho).
\end{aligned}
\end{equation*}
This also shows that every fixed dimension leaves a strictly positive gap to the unrestricted optimum.
\end{proof}

This consequence concerns the optimization itself, independently of any proposed compression map. It rules out an exact finite-dimensional optimizer, but does not rule out finite-dimensional approximations whose dimensions depend on the allowed error. The argument gives no rate of divergence of the required dimension as the error tends to zero.

\section{Compression of sufficiently classical states via the Koashi-Imoto decomposition}
\label{app:KI compression}

The Koashi-Imoto decomposition separates the classical, quantum and redundant parts of the system being compressed \cite{koashi2002}. We consider states that, on the support of $\rho_B$, have the form
\begin{equation}\label{app:eq:KI form}
B=\bigoplus_j Q_j\otimes N_j,
\qquad
\rho_{AB}=\bigoplus_j p_j\rho^{(j)}_{AQ_j}\otimes\omega^{(j)}_{N_j},
\end{equation}
where the $p_j$ are probabilities and the operators $\rho^{(j)}_{AQ_j}$ and $\omega^{(j)}_{N_j}$ are states. The label $j$ can be read without disturbing $\rho_{AB}$. Keeping $j$ and $Q_j$ and discarding $N_j$ gives an exact encoding, since the decoder can prepare $\omega^{(j)}_{N_j}$ again. This redundancy is relative to $A$; an external purification need not be recovered.

For the canonical decomposition, the minimum total dimension for exact local compression is $\sum_{j:p_j>0}|Q_j|$ \cite[Fact 1 and Section III]{Kato2025ExactCompression}. Kato characterizes this dimension without first constructing the decomposition. Approximate compression based on the Koashi-Imoto structure has also been considered in \cite{KuroiwaLeung2023ApproximateKI}. Here we group sectors with similar $AQ$ states to obtain an approximate dimension bound uniform in the number of sectors and the dimensions of the redundant systems. 

\begin{thm}\label{app:thm:KI compression}
Suppose that $\rho_{AB}$ has the form in \cref{app:eq:KI form}, with $|Q_j|\leq r$ for every $j$. For every $0<\epsilon<1$, there exist quantum channels $T:\mathcal B(B)\rightarrow\mathcal B(C)$ and $R:\mathcal B(C)\rightarrow\mathcal B(B)$ such that
\begin{align}
\frac12\left\|\rho_{AB}-(\iden_A\otimes(R\circ T))(\rho_{AB}) \vphantom{2_2^2} \right\|_1&\leq\epsilon,\label{app:eq:KI recovery}\\
|C|&\leq r\left(1+\frac4\epsilon\right)^{|A|^2r^2-1}.\label{app:eq:KI dimension}
\end{align}
The mutual-information loss satisfies
\begin{equation}\label{app:eq:KI information}
0\leq I(A:B)_\rho-I(A:C)_{(\iden_A\otimes T)(\rho)}
\leq2\epsilon\log|A|+g(\epsilon),
\end{equation}
with $g$ as in \cref{lemma:hypothesis are equivalent}. If $A$ is classical in a fixed basis, the exponent in \cref{app:eq:KI dimension} can be replaced by $|A|r^2-1$.
\end{thm}

\begin{proof}
Discard the zero probability sectors. Embed every $Q_j$ isometrically into a common system $Q$ of dimension $r$, using $V_j:Q_j\rightarrow Q$, and put
\begin{equation*}
\tau^{(j)}_{AQ}=(\id_A\otimes V_j)\rho^{(j)}_{AQ_j}(\id_A\otimes V_j^\dagger).
\end{equation*}
The state space on $AQ$ has real affine dimension $|A|^2r^2-1$. A trace-distance $\epsilon/2$-net can be chosen with at most
\begin{equation}\label{app:eq:KI net}
M\leq\left(1+\frac4\epsilon\right)^{|A|^2r^2-1}
\end{equation}
points, by the usual volumetric estimate \cite{AliceBobBanach}. Indeed, the states lie in a ball of radius one for the norm $\frac12\|\cdot\|_1$ on the affine space of trace-one Hermitian operators. A maximal $\epsilon/2$-separated subset is a net, and the disjoint balls of radius $\epsilon/4$ centered at its points lie in a ball of radius $1+\epsilon/4$. Comparing volumes gives \cref{app:eq:KI net}.

Assign each $\tau^{(j)}$ to a nearest net point, with index $c(j)$. States assigned to the same point have trace distance at most $\epsilon$. For each nonempty cell, define
\begin{equation*}\label{app:eq:KI cells}
p_c=\sum_{j:c(j)=c}p_j,
\qquad
\bar\tau^{(c)}_{AQ}=\sum_{j:c(j)=c}\frac{p_j}{p_c}\tau^{(j)}_{AQ}.
\end{equation*}
The encoder $T$ reads $j$, discards $N_j$, applies $V_j$, and keeps the cell index $c(j)$ together with the actual system $Q$. Thus its output on $\rho_{AB}$ is $\bigoplus_c p_c\bar\tau^{(c)}_{AQ}$, with $C$ a direct sum of at most $M$ copies of $Q$. This establishes \cref{app:eq:KI dimension}.

The decoder $R$, on observing $c$, chooses $j$ in that cell with probability $p_j/p_c$, decodes $Q$ into $Q_j$, and prepares $\omega^{(j)}_{N_j}$. A quantum channel that reverses the embedding on its image is
\begin{equation*}\label{app:eq:KI inverse}
D_j(\xi)=V_j^\dagger\xi V_j+\Tr\big[(\id_Q-V_jV_j^\dagger)\xi\big]\zeta_j,
\end{equation*}
where $\zeta_j$ is any state on $Q_j$. Convexity gives $\frac12\|\tau^{(j)}-\bar\tau^{(c(j))}\|_1\leq\epsilon$. Writing $\widehat\rho_{AB}=(\iden_A\otimes(R\circ T))(\rho_{AB})$, addition of trace norms on orthogonal sectors and contractivity under $D_j$ now give
\begin{align*}
\frac12\|\rho_{AB}-\widehat\rho_{AB}\|_1
&=\sum_jp_j\frac12\left\|\rho^{(j)}_{AQ_j}-(\iden_A\otimes D_j)(\bar\tau^{(c(j))}_{AQ})\right\|_1\nonumber\\
&\leq\sum_jp_j\frac12\left\|\tau^{(j)}_{AQ}-\bar\tau^{(c(j))}_{AQ}\right\|_1
\leq\epsilon,
\end{align*}
thus proving \cref{app:eq:KI recovery}. The encoder can be extended arbitrarily to the unused subspace of $B$. Data processing and \cref{lemma:hypothesis are equivalent} with $\delta=2\epsilon$ give \cref{app:eq:KI information}.

If $A$ is classical, all $\tau^{(j)}_{AQ}$ are classical in the same basis of $A$. This state space has affine dimension $|A|r^2-1$, so the same net argument gives the smaller exponent.
\end{proof}

For a uniform bound independent of $|B|$, the dimensions $|Q_j|$ must themselves be controlled independently of $|B|$. In general they can grow with $|B|$, and \cref{app:eq:KI dimension} can then exceed the original dimension. The theorem therefore supplies no such bound for arbitrary states or arbitrary near-optimal auxiliary states. The case $r=1$ includes states classical on the system $B$ being compressed. We remark that a related proof idea has recently been used in~\cite[Corollary 7.5]{Yiwei2026amortized} to give dimension bounds for the amortized channel relative entropy~\cite{wilde2020amortized} under a different restriction. However, while we give compression results per state, \cite{Yiwei2026amortized} bounds the reference dimension needed to approximate the optimal value.

As a corollary of \cref{app:thm:KI compression}, we see that when $|A|=2$ and $\rho_{AB}$ is classical on $A$ with conditional quantum states on $B$ are proportional to projectors, then the desired compression of $B$ is always possible. 

\begin{cor} \label{cor:compression of 2D projectors}
Let $A$ be a quantum system of dimension $|A|=2$ and let $B$ be an arbitrary system. Let $\rho \in \mathcal{B}(AB)$ be any quantum state that is classical on $A$ with conditional quantum states on $B$ that are proportional to projectors. Then, for every $0<\epsilon<1$, there exist quantum channels $T:\mathcal B(B)\rightarrow\mathcal B(C)$ and $R:\mathcal B(C)\rightarrow\mathcal B(B)$ such that
\begin{equation}
\frac12\left\| \vphantom{2-2^2} \rho -(\iden_A\otimes(R\circ T))(\rho )\right\|_1 \leq\epsilon \ , \qquad \  |C| \leq 2 \left(1+\frac4\epsilon\right)^{ 15} . 
\end{equation}
\end{cor}

\begin{proof}
We can w.l.o.g. write $\rho = p \ket{0} \bra{0} \otimes \Pi_1 / \rnk (\Pi_1) + (1-p) \ket{1} \bra{1} \otimes \Pi_2 / \rnk (\Pi_2)$ for some $p \in [0,1]$ and projectors $\Pi_1 , \Pi_2$ on $B$. Jordan's lemma \cite{Jordan1875-vg} (see also \cite[Chap. VII]{Bhatia1996-zd}) states that it is always possible to simultaneously block diagonalize two projectors with each block having dimension $1$ or $2$. Thus, we can act locally on $B$ with a unitary channel $\mathcal{U}_B$ to obtain a state of the form $(\iden_A \otimes \mathcal{U}_B)(\rho) = \bigoplus_j [ (p / \rnk(\Pi_1)) \ket{0} \bra{0} \otimes \Pi_{1,j} + ((1-p) / \rnk (\Pi_2)) \ket{1} \bra{1} \otimes \Pi_{2,j} ] $, where $\Pi_{1,j}$ and $\Pi_{2,j}$ are components of $\Pi_1=\bigoplus_j \Pi_{1,j}$ and $\Pi_2 = \bigoplus_j \Pi_{2,j}$ in the $j$'th block of the simultaneous block diagonal structure. By redistributing some constants, it can be seen that this state is of the form from \cref{app:eq:KI form} with $N_j$ trivial and $Q_j$ corresponding to the space spanned by the $j$'th block. Thus, using the result from \cref{app:thm:KI compression} with $ |Q_j| \leq 2 \coloneqq r$ after applying the unitary channel $\mathcal{U}_B$ proves the result. 
\end{proof}

\subsection{Strong compression for classical systems}
\label{app:KI classical cases}
\label{app:classical_compression_results}

A special case of the compression described above is when the system to be compressed $B$ is completely classical, in which case \cref{app:thm:KI compression} entails that the strong compression described in the Introduction is possible. However, due to the significance of this case, which also includes the completely classical analogue of the problem discussed in this Letter, we here investigate this case further and prove stronger compression results that also improve upon earlier known results, see \cref{app:eq:classical dimension comparison}. We here consider states of the form
\begin{equation}\label{app:eq:classical B form}
\rho_{AB}=\sum_jp_j\sigma_{A,j}\otimes\ket j\!\bra j_B.
\end{equation}

\begin{cor}
\label{app:cor:classical special cases}
\label{prop:compression probability distributions}
\label{prop:compression QC states}
Consider any two quantum systems $A$ and $B$, with $|A|\geq2$, and any bipartite state $\rho_{AB}$ of the form in \cref{app:eq:classical B form}. Then, for every $0<\epsilon<1$, there exists a further system $C$, whose size depends only on $A$ and $\epsilon$, and a pair of quantum channels $T:\mathcal B(B)\rightarrow\mathcal B(C)$ and $R:\mathcal B(C)\rightarrow\mathcal B(B)$ satisfying \cref{app:eq:KI recovery}, with
\begin{align}
|C|&\leq\left(1+\frac4\epsilon\right)^{|A|^2-1},\label{app:eq:classical B dimension}\\
0\leq I(A:B)_\rho-I(A:C)_{(\iden_A\otimes T)(\rho)}
&\leq\frac\epsilon2\log|A|+g (\epsilon/2).\label{app:eq:classical B information}
\end{align}
The channels act classically on $B$ and $C$, and the recovered state has the original marginal $\rho_B$.

If $A$ is also classical, this gives a pair of classical channels $T:Y\rightarrow Z$ and $R:Z\rightarrow Y$ for any joint probability distribution $p_{XY}$, identifying $A=X$, $B=Y$ and $C=Z$. In this case, one can further choose
\begin{equation}\label{app:eq:classical grid dimension}
|C|\leq\binom{\lceil|A|/\epsilon\rceil+|A|-1}{|A|-1}
\leq \left(\frac4\epsilon\right)^{|A|-1}
\end{equation}
with the same recovery and information-loss guarantees. The case $|A|=1$ is trivial.
\end{cor}

\begin{proof}
We first choose a trace-distance $\epsilon/2$-net $\mathcal N_{\epsilon/2}$ on the states of $A$, and for each $j$ choose $\omega_j\in\mathcal N_{\epsilon/2}$ with $\frac12\|\sigma_{A,j}-\omega_j\|_1\leq\epsilon/2$. Label the net points by $c$, and let $c(j)$ be the label of $\omega_j$. We can now define the channels explicitly as
\begin{equation*}\label{app:eq:classical channels}
\begin{aligned}
T(X)&=\sum_j\bra jX\ket j\,\ket{c(j)}\!\bra{c(j)},\\
R(Y)&=\sum_{c:p_c>0}\bra cY\ket c\sum_{j:c(j)=c}\frac{p_j}{p_c}\ket j\!\bra j,
\qquad p_c\coloneqq\sum_{j:c(j)=c}p_j.
\end{aligned}
\end{equation*}
Only cells with positive probability are needed; the encoder is extended arbitrarily on zero-probability input symbols. The encoder groups those conditional states that are close to the same net point, and the decoder reconstructs the original marginal on $B$ by resampling $j$ within its cell. Thus, only conditional states that are close to each other get mixed. These are precisely the channels in \cref{app:thm:KI compression} with one-dimensional quantum and redundant parts, which proves \cref{app:eq:KI recovery,app:eq:classical B dimension}. In particular, the rounded state
\begin{equation*}
\label{app:eq:classical rounded state}
\eta_{AB}\coloneqq\sum_jp_j\omega_j\otimes\ket j\!\bra j
\end{equation*}
satisfies $\frac12\|\rho_{AB}-\eta_{AB}\|_1\leq\epsilon/2$ and $(\iden_A\otimes(R\circ T))(\eta_{AB})=\eta_{AB}$. The reconstruction preserves $\rho_B$ exactly.

For fully classical states we can construct the net explicitly. Define $M\coloneqq\lceil|A|/\epsilon\rceil$ and
\begin{equation*}\label{app:eq:classical grid net}
\mathcal N_{\epsilon/2}\coloneqq\left\{\left(\frac{m_1}{M},\ldots,\frac{m_{|A|}}{M}\right)
\;\middle|\;m_j\in\mathbb N_0,\ \sum_jm_j=M\right\}.
\end{equation*}
Round the coordinates of $Mp_X$ down and then round up enough coordinates to make their sum $M$. Each coordinate changes by at most $1/M$, so this is a trace-distance $\epsilon/2$-net. The number of grid points is $\binom{M+|A|-1}{|A|-1}$. Comparing the factors in the binomial product gives \cref{app:eq:classical grid dimension}. We use the same channels as above, with the conditional distributions $p(X|y)$ in place of $\sigma_{A,j}$.

It remains to improve the mutual-information estimate. In each cell put $\bar\sigma_A^{(c)}=\sum_{j:c(j)=c}(p_j/p_c)\sigma_{A,j}$. The information lost by retaining only $c(j)$ is
\begin{equation*}\label{app:eq:classical cell Holevo}
I(A:B)_\rho-I(A:C)_{(\iden_A\otimes T)(\rho)}
=\sum_cp_c\left[H(\bar\sigma_A^{(c)})-\sum_{j:c(j)=c}\frac{p_j}{p_c}H(\sigma_{A,j})\right].
\end{equation*}
Each bracket is the Holevo quantity of the conditional ensemble in that cell. Its average trace distance from the corresponding net point is at most $\epsilon/2$. Shirokov's reference-state bound \cite{Shirokov2019HolevoBounds} bounds each bracket by $(\epsilon/2)\log|A|+g(\epsilon/2)$. Averaging proves \cref{app:eq:classical B information}, including the case of diagonal (i.e. classical) states and the grid covering.

The extension from probability distributions to general states of the form \eqref{app:eq:classical B form} requires only a net on the states of $A$, and the direct covering argument in real affine dimension $|A|^2-1$ from \cref{app:thm:KI compression} gives such a net.
\end{proof}
For comparison with \cite[Eqs.~(5) and (19)]{Hirche2020-np}, we fix the trace-distance recovery error at $\epsilon$ throughout. Their net error is then $\epsilon/2$, and the dimension bounds read
\begin{equation}\label{app:eq:classical dimension comparison}
\begin{array}{c|c|c}
&\text{Ref.~\cite{Hirche2020-np}}&\text{Here}\\ \hline
A\text{ classical}
&(4/\epsilon)^{|A|}
&(4/\epsilon)^{|A|-1}\\[8pt]
A\text{ quantum}
&(1+4/\epsilon)^{2|A|^2}
&(1+4/\epsilon)^{|A|^2-1}.
\end{array}
\end{equation}
In both rows $B$ is classical. At the same recovery error, \cref{app:eq:classical B information} also improves upon the information-loss guarantee of $\epsilon\log|A|+g(\epsilon)$ from \cite[Cors.~4 and 9]{Hirche2020-np}.

\section{Exact bounds for restricted auxiliary optimizations}
\label{app:restricted optimizations}

We now apply the exact encoding from \cref{app:KI compression} to the optimizations in \cref{app:bottleneck squashed}. Fix a positive integer $r$. In both cases, we require the marginal $\omega_{AC}$ to have the form in \cref{app:eq:KI form}, with $B$ replaced by $C$ and all $|Q_j|\leq r$. There is no bound on the number of sectors or the dimensions of the redundant systems. Only the $AC$ marginal is restricted; the full $AE$ marginal remains fixed. 

We make use the following elementary classical support argument.
\begin{lem}\label{app:lem:restricted support}
Let $\tau_j$ be finitely many states on a fixed system $F$, with probabilities $p_j$, real numbers $f_{ij}$ ($1\leq i\leq k$), and real costs $c_j$. There are probabilities $q_j$ supported on at most $|F|^2+k$ of the same branches, preserving the average state and the $k$ scalar averages, with $\sum_jq_jc_j\leq\sum_jp_jc_j$.
\end{lem}
\begin{proof}
Minimize the linear cost over the compact polytope of weights with the prescribed averages, and choose an extreme-point minimizer. There are at most $|F|^2+k$ independent real equalities, including normalization. A larger support would admit a nonzero feasible perturbation in both signs, contradicting extremality.
\end{proof}

\begin{prp}\label{app:prp:restricted QIB}
For every eligible channel $T$ in \cref{app:eq:QIB definition} whose $AC$ marginal has quantum parts of dimension at most $r$, there is another channel $\widetilde T:\mathcal B(B)\rightarrow\mathcal B(C')$, with output $\widetilde\omega$, such that
\begin{equation}
\begin{aligned}
|C'|&\leq r\big[(\rnk\rho_B)^2+1\big],\label{app:eq:restricted QIB dimension}\\
I(A:C')_{\widetilde\omega}&=I(A:C)_\omega,
\\
I(AE:C')_{\widetilde\omega}&\leq I(AE:C)_\omega.
\end{aligned}
\end{equation}
The new $AC'$ marginal also has quantum parts of dimension at most $r$. For every nonnegative rate, the supremum restricted to such channels is attained in the dimension in \cref{app:eq:restricted QIB dimension}.
\end{prp}
\begin{proof}
Apply the exact encoder associated with the decomposition of $\omega_{AC}$: read the sector $j$, discard its redundant system, and embed its quantum part into a common system $Q$ with $|Q|=r$. The resulting state is
\begin{equation}\label{app:eq:restricted flagged state}
\gamma_{AEJQ}=\sum_jp_j\ket j\!\bra j_J\otimes\gamma^{(j)}_{AEQ}.
\end{equation}
Its $AE$ marginal is unchanged and has rank $\rnk\rho_B$. Exact recovery of $AC$ and data processing give $I(A:JQ)_\gamma=I(A:C)_\omega$ and $I(AE:JQ)_\gamma\leq I(AE:C)_\omega$.

Every positive-weight $\gamma^{(j)}_{AE}$ is supported on the support of this fixed marginal. Apply \cref{app:lem:restricted support} on that support, preserving the scalar $H(A|Q)_{\gamma^{(j)}}$ and minimizing the cost $-H(AE|Q)_{\gamma^{(j)}}$. At most $(\rnk\rho_B)^2+1$ branches suffice. Since
\begin{align*}
I(A:JQ)_\gamma&=H(A)_\gamma-\sum_jp_jH(A|Q)_{\gamma^{(j)}},\\
I(AE:JQ)_\gamma&=H(AE)_\gamma-\sum_jp_jH(AE|Q)_{\gamma^{(j)}},
\end{align*}
the reweighted state preserves relevance and does not increase the cost. Keep the remaining label $J'$ and $Q$ as $C'$. Every extension of the fixed $AE$ marginal can be generated by a channel on its purifier $B$, so this state defines $\widetilde T$. The reweighting need not be a postprocessing of the original $C$; it replaces the encoder on $B$.

For attainment, fix $|J'|=(\rnk\rho_B)^2+1$ and $|Q|=r$. States with the fixed $AE$ marginal, classical on $J'$, and satisfying the rate constraint form a compact set. Every such state has the required $AC'$ structure, and the reduction loses no restricted optimum. Continuity of mutual information therefore gives a maximum.
\end{proof}

\begin{prp}\label{app:prp:restricted squashed}
For every extension $\omega_{AEC}$ of a state $\sigma_{AE}$ whose $AC$ marginal has quantum parts of dimension at most $r$, there is another extension $\widetilde\omega_{AEC'}$ of the same state such that
\begin{equation}
\begin{aligned}
|C'|&\leq r(\rnk\sigma_{AE})^2,\label{app:eq:restricted squashed dimension}\\
\frac12 I(A:E|C')_{\widetilde\omega}&\leq\frac12 I(A:E|C)_\omega.
\end{aligned}
\end{equation}
The new $AC'$ marginal has quantum parts of dimension at most $r$. The infimum over extensions with this restriction is attained in the dimension in \cref{app:eq:restricted squashed dimension}.
\end{prp}
\begin{proof}
Use the same exact encoder on $C$, giving an extension $\gamma$ of $\sigma_{AE}$ of the form in \cref{app:eq:restricted flagged state}. Exact recovery of $AC$ preserves $I(A:C)$, while data processing cannot increase $I(A:EC)$. Hence
\begin{align*}
I(A:E|JQ)_\gamma
&=I(A:EJQ)_\gamma-I(A:JQ)_\gamma\nonumber\\
&\leq I(A:EC)_\omega-I(A:C)_\omega
=I(A:E|C)_\omega.
\end{align*}
The left hand side is $\sum_jp_jI(A:E|Q)_{\gamma^{(j)}}$. Each positive-weight $\gamma^{(j)}_{AE}$ lies on the support of $\sigma_{AE}$. Apply \cref{app:lem:restricted support} on this support with no additional scalar constraint and costs $I(A:E|Q)_{\gamma^{(j)}}$. This preserves $\sigma_{AE}$ and does not increase the conditional mutual information, using at most $(\rnk\sigma_{AE})^2$ branches. Keeping these branches and $Q$ proves the bounds. The minimum exists by compactness and continuity on the set of extensions with fixed $\sigma_{AE}$, classical on a fixed $J'$ of this size, and $|Q|=r$, as in the preceding proof.
\end{proof}

These statements give exact bounds for the restricted optimizations. If a near-optimal unrestricted candidate has quantum parts bounded by $r$, it can therefore be replaced at the same accuracy by one in the displayed dimension. The dimension hypothesis is essential, as the binary example shows.

\begin{cor}\label{app:cor:restricted quantum gap}
For the dephased Bell state $\sigma^{(c)}_{AE}$ in \cref{app:eq:dephased Bell}, every extension with $AC$ quantum parts of dimension at most $r$ satisfies
\begin{equation}\label{app:eq:restricted squashed gap}
\frac12 I(A:E|C)_\omega\geq E_{4r}(\sigma^{(c)}_{AE})>E_{\mathrm{sq}}(\sigma^{(c)}_{AE}).
\end{equation}
For the binary source in \cref{app:eq:binary source}, every eligible channel at rate $H(B)_\rho$ with this restriction satisfies
\begin{equation}\label{app:eq:restricted QIB gap}
I(A:C)_\omega\leq1-E_{4r}(\sigma^{(c)}_{AE})<\mathsf I_\rho(H(B)_\rho).
\end{equation}
Thus the dimensions of the quantum parts must diverge along any sequence approaching either optimum, even with unrestricted finite classical storage and redundant systems.
\end{cor}
\begin{proof}
The state $\sigma^{(c)}_{AE}$ has rank two, so \cref{app:prp:restricted squashed} bounds its restricted infimum from below by $E_{4r}$. The strict decrease of $E_D$ used in \cref{app:cor:binary nonattainment} gives $E_{4r}>E_{\mathrm{sq}}$. At rate $H(B)_\rho$, \cref{app:eq:bottleneck entropy identity} gives $I(A:C)_\omega\leq1-I(A:E|C)_\omega/2$. Combining this with \cref{app:eq:bottleneck identity,app:eq:restricted squashed gap} proves \cref{app:eq:restricted QIB gap}.
\end{proof}

\end{appendices}

\end{document}